\documentclass[preprint,showpacs,preprintnumbers,amsmath,amssymb]{revtex4-2}
\usepackage{graphicx}
\usepackage{epsfig}
\usepackage{bm}
\usepackage{amsfonts}
\usepackage{subfigure}
\usepackage{multirow}
\usepackage{float}
\usepackage{color}
\usepackage[colorlinks]{hyperref}

\begin{document}

\title{\textbf{Primordial black holes ensued from exponential potential and coupling parameter in nonminimal derivative inflation model}}

%\title{\textbf{Generation of primordial black holes and induced gravitational waves from quartic potential  in nonminimal derivative coupling %inflationary model}}

\author{Soma Heydari\footnote{s.heydari@uok.ac.ir} and Kayoomars Karami\footnote{kkarami@uok.ac.ir}}
\address{\small{Department of Physics, University of Kurdistan, Pasdaran Street, P.O. Box 66177-15175, Sanandaj, Iran}}
\date{\today}

%============================================Abstract===================================================
\begin{abstract}

Here, Primordial Black Holes (PBHs) creation from exponential potential has been inquired, through gravitationally raised friction  emanated from the nonminimal coupling between gravity and field derivative setup.  Setting a two-parted exponential function of inflaton field as coupling parameter, and fine-tuning of four parameter Cases of our model, we could sufficiently slow down the inflaton owing to high friction during an ultra slow-roll phase. This empowers us to achieve enough enhancement in the amplitude of curvature perturbations power spectra, via numerical solving of Mukhanov-Sasaki equation. Thereafter, we illustrate the generation of four PBHs with disparate masses in RD era, corresponding to our four parameter Cases. Two specimens of these PBHs with stellar ${\cal O}(10)M_{\odot}$ and earth  ${\cal O}(10^{-6})M_{\odot}$ masses can be appropriate to explicate the LIGO-VIRGO events, and  the ultrashort-timescale microlensing events in OGLE data, respectively. Another two Cases of PBHs have asteroid masses around ${\cal O}(10^{-13})M_{\odot}$ and ${\cal O}(10^{-15})M_{\odot}$  with abundance of $96\%$ and $95\%$  of the Dark Matter (DM) content of the universe. Furthermore, we scrutinize  the induced Gravitational Waves (GWs) ensued from PBHs production in our model. Subsequently, we elucidate that their contemporary density parameter spectra $(\Omega_{\rm GW_0})$ for all predicted Cases  have acmes which lie in the sensitivity scopes of the GWs detectors, thereupon  the verity of our conclusions can be verified in view of deduced data from these detectors.  At length, our numerical outcomes exhibit a power-law behavior for the spectra of $\Omega_{\rm GW_0}$ with respect to frequency as $\Omega_{\rm GW_0} (f) \sim (f/f_c)^{n} $ in the proximity of acmes position. As well, in the infrared regime $f\ll f_{c}$, the log-reliant form of power index as $n=3-2/\ln(f_c/f)$ is attained.
 \end{abstract}
\maketitle

%============================================Introduction===================================================
\newpage
\section{Introduction}
It is known that, enough sizable amplitude of curvature perturbations in the inflationary epoch could give rise to form the ultra-condensed districts of the primal cosmos, thence gravitationally collapse of these sectors terminate in generation of Primordial Black Holes (PBHs) in the Radiation Dominated (RD) era. The first intimation of this outline should be addressed to the researches   \cite{zeldovich:1967,Hawking:1971,Carr:1974,carr:1975}.\\
The enigmatic nature of Dark Matter (DM)  \cite{Garrett:2011}, and unsuccessful observation of particle DM, beside the latest thriving revelation  of  Gravitational Waves (GWs) emerged from the coalescence of two black holes  with masses around  $ 30 M_\odot$ ($M_\odot$ signifies the solar mass) by LIGO-Virgo Collaboration \cite{Abbott:2016-a,Abbott:2016-b,Abbott:2017-a,Abbott:2017-b,Abbott:2017-c}, have widely inspired the researchers  to contemplate PBHs as interesting source for the entire or a percentage of the universe DM content and GWs \cite{Ivanov:2017,Frampton:2010,Inomata:2017,Carr:2020,Bird:2016,Clesse:2017,Sasaki:2016,Carr:2016,Belotsky:2014,
Belotsky:2019,OGLE,HSC,Katz,Khlopov:2010,Cai:2018,Ballesteros:2020a,Kamenshchik:2019,fu:2019,Dalianis:2019,
mahbub:2020,mishra:2020,Fumagalli:2020a,Braglia:2020,Braglia:2021,Dalianis:2021,Garcia-Bellido:2017pdu,Germani:2017,
Ballesteros:2018,Liu:2020,Solbi-a:2021,Solbi-b:2021,Teimoori-b:2021,Laha:2019,Teimoori:2021,
fu:2020,Dalianis:2020,Dasgupta:2020,Pi,Garcia-Bellido:2017,Motohashi:2017,Heydari:2021,Rezazadeh:2021,Yogesh:2021,
Domenech-a:2020,Domenech-b:2021,Kimura:2021,Kawai:2021,Kawai:2021edk,Lin:2020,Lin-b:2021,Lu:2020,Zhang:2021,Domenech:2020b}.
By reason of the non-stellar inception of PBHs generation, their masses are not confined to Chandrasekhar restriction and they could place in the vast range of masses. PBHs at the  mass scope of  ${\cal O}(10^{-5})M_\odot$   with the fractional exuberance around ${\cal O}(10^{-2})$, can be regarded as the genesis of ultrashort-timescale microlensing events in the OGLE data, because of settling  in the sanctioned area by OGLE data \cite{OGLE}. Furthermore PBHs within the scope of asteroid masses ${\cal O}(10^{-16}-10^{-11})M_\odot$ can comprise all DM content of the universe  \cite{Laha:2019,Katz,HSC,Solbi-a:2021,Teimoori:2021,Teimoori-b:2021,Solbi-b:2021,Dasgupta:2020,Heydari:2021}, inasmuch as the gravitational femtolensing of gamma-ray bursts \cite{Barnacka} is inoperative by way of weakening the lensing  effects via the wave effects \cite{HSC,Katz}, and Subaru Hyper Supreme-Cam (Subaru HSC) microlensing observations decree no constraint on PBHs in the mass ranges lower than $10^{-11}M_{\odot}$. Moreover, the imposed constraint by white dwarf \cite{WD} has been shown to be ineffective using numerical simulations in \cite{Montero-Camacho}.

It is well understood that, so as to generate detectable PBHs in RD stage, the amplitude of primordial curvature perturbations (${\cal R}$) must be amplified to specific order during inflationary era. Recrudescence of
superhorizon scales  related to enhanced amplitude of ${\cal R}$  to the horizon in RD domain, gives rise to formation of ultradense zones, and PBHs can be produced from  gravitationally collapse of these zones. A sufficient amplification of the power spectrum of ${\cal R}$ to order  ${\cal P}_{\cal R}\sim{\cal O}(10^{-2})$ at small scales is necessitated to generate detectable PBHs, whereas the recent observations of CMB anisotropies confined the power spectrum of ${\cal R}$ to ${\cal P}^{*}_{\cal R}\sim2.1 \times 10^{-9}$ \cite{akrami:2018} at pivot scale $k_{*}=0.05~ \rm Mpc^{-1}$.
Heretofore, multifarious technical methods have been suggested  by researchers to achieve  amplified  curvature power spectrum with the amount around  $10^{7}$ times larger  than scalar power spectrum at CMB scales \cite{Ivanov:2017,Frampton:2010,Inomata:2017,Carr:2020,Bird:2016,Clesse:2017,Sasaki:2016,Carr:2016,Belotsky:2014,
Belotsky:2019,OGLE,HSC,Katz,Khlopov:2010,Cai:2018,Ballesteros:2020a,Kamenshchik:2019,fu:2019,Dalianis:2019,
mahbub:2020,mishra:2020,Fumagalli:2020a,Braglia:2020,Braglia:2021,Dalianis:2021,Garcia-Bellido:2017pdu,Germani:2017,
Ballesteros:2018,Liu:2020,Solbi-a:2021,Solbi-b:2021,Teimoori-b:2021,Laha:2019,Teimoori:2021,
fu:2020,Dalianis:2020,Dasgupta:2020,Pi,Garcia-Bellido:2017,Motohashi:2017,Heydari:2021,Rezazadeh:2021,Yogesh:2021,
Domenech-a:2020,Domenech-b:2021,Kimura:2021,Kawai:2021,Kawai:2021edk,Lin:2020,Lin-b:2021,Lu:2020,Zhang:2021,
Domenech:2020b}. Pi \emph{et al}. in \cite{Pi} could attain this amplification in ${\cal P}_{\cal R}$  using the Starobinsky $R^2$ model with a non-minimally coupled scalar field $\chi$, through the mortal oscillations during the transition from the first phase of inflation to the second one. Recently, a similar model of two-field inflation has been proposed by Braglia \emph{et al}.  for generating  PBHs and GWs \cite{Braglia:2020}. In the model unlike that of Pi \emph{et al}. \cite{Braglia:2021} a very broad peak in the Scalar power spectrum and therefore PBHs with a broader mass function, and also specific oscillatory pattern in the stochastic gravitational wave background, akin to  \cite{Barnacka}, could have been produced. Further similar model for producing oscillatory GWs is presented in \cite{Dalianis:2021}.  Another method to amplify ${\cal P}_{\cal R}$ is proposed by \cite{Dalianis:2019,mahbub:2020,Garcia-Bellido:2017,Motohashi:2017,Garcia-Bellido:2017pdu,Germani:2017,Ballesteros:2018} through applying inflationary potentials for their models with an inflection point. In these models an enhancement in the scalar power spectrum is attained by slowing down the inflaton field during a transient era of Ultra Slow-Roll (USR) inflation in comparison with Slow-Roll inflation because of an inflection point in the potential. Another appropriate way to produce an USR phase and slowing down the inflaton is increasing the friction by applying the NonMinimal Derivative Coupling to gravity (NMDC) framework  for the  models
\cite{Germeni:2010,Defelice:2011,Tsujikawa:2012,Tsujikawa:2013,Defelice:2013,Dalianis:2020,fu:2019,fu:2020,
Teimoori:2021,Heydari:2021}.

The framework of NMDC constructed of nonminimal coupling between field derivative and the Einstein tensor is a subdivision of the comprehensive  scalar-tensor theories with second order equation of motion, like as General Relativity (GR), to wit the Horndeski theory \cite{Defelice:2011,Tsujikawa:2012,Tsujikawa:2013,Defelice:2013,Horndeski:1974}. The Horndeski theory forestalls the model from negative energy and the Ostrogradski instability \cite{ostrogradski:1850,Langlois,Chen,Rham}.

It is known that, exponential potential drives an endless inflationary epoch in standard model of inflation \cite{karami:2017}. Moreover,
as regards the inconsistency of the exponential potential in the standard framework with  Planck  2018  TT,TE,EE+lowE+lensing+BK14+BAO data at CMB scale \cite{akrami:2018,karami:2017}, we try to reclaim this potential in NMDC setup. In other words the feature of gravitationally enhanced friction in NMDC setup inspires us to evaluate the compatibility of exponential form of potential in this framework with the latest Planck's observational data on large scales, and furthermore  examine the likelihood of PBHs generation in detectable  mass scopes at small scales contemporaneously. In \cite{fu:2019} the production of PBHs in transient NMDC framework with a power-law potential $V(\phi)\propto \phi^{2/5}$ has been investigated by Fu \emph{et al}. but in their work the NMDC term just operate within the USR epoch and dose not have an indelible impression due to the selected form of coupling parameter between field derivative and gravity. In the present work, we have selected a special exponential form of two-parted coupling function so as to
have the NMDC framework  all over the inflationary era.

Dissemination of produced secondary gravitational waves coeval with PBHs generation, could be the further upshot of reverting the enhanced amplitude of primordial scalar perturbations to the horizon in RD era
\cite{Kohri:2018,Cai:2019-a,Cai:2019-b,Bartolo:2019-a,Bartolo:2019-b,Wang:2019,Cai:2019-c,Xu:2020,Lu:2019,Hajkarim:2019,
Domenech-a:2020,Domenech:2020b,Fumagalli:2020b,Motohashi:2017,Germani:2017,Di:2018,Namba:2015,Garcia-Bellido:2017, Lu:2019,Kawasaki:2016,Kannike:2017,Garcia-Bellido:1996,Clesse:2015,Teimoori:2021,Teimoori-b:2021,Solbi-b:2021,Solbi-a:2021,
Heydari:2021}.  In this article, we  compute the present density parameter spectra of induced GWs as a supplementary effect of PBHs generation in  NMDC framework, and peruse the verity of our numerical outcomes in comparison with  the  sensitivity scopes of  multifarious GWs detectors. At length we appraise the inclination of the energy spectra of GWs in disparate ranges of frequency.

This paper is classified  as  follows. In Sec. \ref{sec2}, we review succinctly  the foundation of   nonminimal derivative coupling structure. The  Sec. \ref{sec3}, is given over to clarifying  the appropriate technique to amplify the amplitude of the curvature power spectrum at small scale to  order ${\cal O}(10^{-2})$ in NMDC model. Thence, we investigate the feasibility of PBHs generation with varietal  masses and fractional abundances in Sec. \ref{sec5}, and present energy spectra of induced GWs in our setup are computed in Sec. \ref{sec6}. After all, the foreshortened outcomes are enumerated in Sec. \ref{sec7}.
%==========================================NMDC model======================================================
\section{Nonminimal Derivative Coupling framework}\label{sec2}
The NonMinimal Derivative Coupling (NMDC) model is depicted by the generic action \cite{Germeni:2010,Defelice:2011,Tsujikawa:2012,Tsujikawa:2013,Defelice:2013} as
\begin{equation}\label{action}
S=  \int {\rm d}^{4}x\sqrt{-g}\bigg[\frac{1}{2}R-\frac{1}{2}\big(g^{\mu\nu}-\xi G^{\mu\nu}\big)\partial_{\mu}\phi\partial_{\nu}\phi-V(\phi)\bigg],
\end{equation}
wherein the derivative of inflaton field is coupled to the Einstein tensor via the coupling parameter denoted by  $\xi$ with dimension of  $({\rm  mass})^{-2}$, and  $g$ is  determinant of the metric tensor $ g_{{\mu}{\nu}}$, $R$ is the Ricci scalar, $G^{{\mu}{\nu}}$ is the Einstein tensor,  and $V(\phi)$ signifies the potential of the scalar field $\phi$. Referring to our erstwhile explanation, this action (\ref{action}) pertains to general scalar-tensor theories viz the Horndeski theory with the second order equations of motion.
The general Lagrangian of this theories embodies the expression $G_{5}(\phi,X)G^{\mu\nu}(\nabla_{\mu}\nabla_{\nu}\phi)$, wherein  $G_{5}$ is a generic function of $\phi$ and kinetic term $X=-\frac{1}{2}g^{\mu\nu}\partial_{\mu}\phi \partial_{\nu}\phi$. Presuming  $G_{5}=-\Theta(\phi)/2$, and $\xi\equiv d\Theta/d\phi$, thereafter integrating partially, the NMDC action  (\ref{action}) is retrieved from the Horndeski Lagrangian. The coupling parameter $\xi$ can be considered as a constant parameter \cite{Germeni:2010,Defelice:2011,Tsujikawa:2012,Tsujikawa:2013,Defelice:2013}, or as a function of $\phi$ \cite{Granda:2020,Teimoori:2021,Dalianis:2020,Heydari:2021}.

With this in mind, we excogitate $\xi=\theta(\phi)$ as an exponential two-parted function of $\phi$, so as to not only ameliorate the observational prognostications of the exponential  potential on large scales, but also generation of  PBHs and induced GWs on small scales  could be expounded prosperously in NMDC setup. At first we initiate to study the dynamics of  homogeneous and isotropic background having the
flat Friedmann-Robertson-Walker (FRW) metric as $g_{{\mu}{\nu}}={\rm diag}\Big(-1, a^{2}(t), a^{2}(t), a^{2}(t)\Big)$,
in which  $a(t)$ and $t$  denote the scale factor and cosmic time.
Thenceforth, we reconsider the attained equations for propagated  perturbations during inflation era in NMDC model depicted through action (\ref{action}).

The Friedmann equations and the equation of motion governing the scalar field $\phi$ ensued from taking derivative of action (\ref{action}) with regard to $g_{\mu\nu}$ and $\phi$ can be obtained as following form
\begin{align}
& 3H^{2}-\frac{1}{2}\Big(1+9H^{2}\theta(\phi)\Big)\dot{\phi}^{2}-V(\phi)=0,
\label{FR1:eq}
\\
& 2\dot{H}+\left(-\theta(\phi)\dot{H}+3\theta(\phi)H^{2}+1\right)\dot{\phi}^{2}-H\theta_{,\phi}\dot{\phi}^{3}
-2H\theta(\phi)\dot{\phi}\ddot{\phi}=0,
\label{FR2:eq}
\\
& \left(1+3\theta(\phi)H^{2}\right)\ddot{\phi}+\Big(1+\theta(\phi)(2\dot{H}+3H^{2})\Big)3H\dot{\phi}
 +\frac{3}{2}\theta_{,\phi}H^{2}\dot{\phi}^{2}+V_{,\phi}=0,
\label{Field:eq}
\end{align}
where $H\equiv \dot{a}/a $ signifies  the Hubble parameter,  the dot symbol implies derivative with regard to the cosmic time $t$, and $({,\phi})$ denotes derivative with regard to $\phi$.
We also stipulate that the reduced Planck mass equates with one $(M_P=1/\sqrt{8\pi G}=1)$, all over this article.
Pursuant the calculations of \cite{Defelice:2011,Tsujikawa:2012} in NMDC framework, slow-roll parameters are acquainted as follows
\begin{equation}\label{SRP}
  \varepsilon \equiv -\frac{\dot H}{H^2}, \hspace{.5cm}  \delta_{\phi}\equiv \frac{\ddot{\phi}}{ H\, \dot{\phi}}, \hspace{.5cm}\delta_{X}\equiv \frac{\dot{\phi}^2}{2 H^2}, \hspace{.5cm} \delta_{D}\equiv \frac{\theta(\phi)\dot{\phi}^2}{4}.
\end{equation}
The slow-roll approximation of inflation is  validated provided that $\{\epsilon, |\delta_{\phi}|, \delta_{X},\delta_{D}\}\ll 1$, and thereunder the potential energy term of energy density is prevailing owing to negligibility of kinetic energy term. Thus the equations (\ref{FR1:eq})-(\ref{Field:eq}) can be recast as the following abridged form
\begin{align}
\label{FR1:SR}
& 3 H^2\simeq V(\phi),\\
  \label{FR2:SR}
& 2\dot{H}+{\cal A}\dot{\phi}^2-H \theta_{,\phi}\dot{\phi}^3\simeq0,\\
  \label{Field:SR}
& 3 H\dot{\phi}{\cal A}+\frac{3}{2}\theta_{,\phi}H^2\dot{\phi}^2+V_{,\phi}\simeq0,
\end{align}
in which
\begin{equation}\label{A}
{\cal A}\equiv 1+3 \theta(\phi) H^2.
\end{equation}
On the presumption that the following condition
\begin{equation}\label{condition}
|\theta_{,\phi}H\dot{\phi}|\ll {\cal A},
\end{equation}
is confirmed during slow-roll epoch, the equations (\ref{FR1:SR})-(\ref{Field:SR}) can be simplified as
\begin{align}
\label{FR1:SRC}
& 3 H^2\simeq V(\phi),\\
  \label{FR2:SRC}
& 2\dot{H}+{\cal A}\dot{\phi}^2\simeq0,\\
  \label{Field:SRC}
& 3 H\dot{\phi}{\cal A}+V_{,\phi}\simeq0.
\end{align}
Utilizing  equations (\ref{FR1:SRC}) and (\ref{Field:SRC}), the first slow-roll parameter can be written as
\begin{equation}\label{epsilon}
\varepsilon \simeq \delta_{X}+6\delta_{D}\simeq \frac{\varepsilon_{V}}{{\cal A}},
\end{equation}
in which
\begin{equation}\label{epsilonv}
\varepsilon_{V}\equiv \frac{1}{2}\left(\frac{V_{,\phi}}{V}\right)^2.
\end{equation}
It is clear from  (\ref{epsilon}) that, ${\cal A}\simeq 1$ leads to  $\varepsilon\simeq \varepsilon_{V}$, and the standard model of  slow-roll inflation is retrieved. Furthermore, ${\cal A}\gg 1$ gives rise to $\varepsilon \ll\varepsilon_{V}$ and considerable increment in the friction, and as a consequence more slowing down the inflaton field during USR phase with regard the slow-roll outlook. Severe lessening of the velocity of inflaton field in USR domain due to enlarged friction results in acute reduction of the first slow-roll parameter and thence remarkable increment in the scalar power spectrum. Hereupon, we ponder the power spectrum of ${\cal R}$ in NMDC setup at the instant of Hubble horizon traversing via comoving wavenumber $k$ \cite{Tsujikawa:2013} as follows
\begin{equation}\label{Ps}
{\cal P}_{\cal R}=\frac{H^2}{8 \pi ^{2}Q_{s}c_{s}^3}\Big|_{c_{s}k=aH}\,,
\end{equation}
where, in accordance with computations of \cite{Tsujikawa:2012}, we have
\begin{align}\label{Qs,cs2}
& Q_{s}= \frac{ w_1(4 w_{1}{w}_{3}+9{w}_{2}^2)}{3{w}_{2}^2},\\
 & c_{s}^2=\frac{3(2{w}_{1}^2 {w}_2 H-{w}_{2}^2{w}_4+4{w}_1\dot{w_1}{w}_2-2{w}_{1}^2\dot{w_2})}{{w}_1(4{w}_{1}{w}_{3}+9{ w}_{2}^2)},
\end{align}
and
\begin{align}
\label{w1}
& {w}_1=1-2\delta_D,\\
  \label{W2}
& {w}_2=2H(1-6\delta_D),\\
& {w}_3=-3 H^2(3-\delta_{X}-36\delta_D),\\
&{w}_4=1+2\delta_D.
\end{align}
Utilizing equations (11)-(13) associated with the  background evolution
under  the slow-roll approximation, the scalar power spectrum (\ref{Ps}) takes the following form
\begin{equation}\label{PsSR}
{\cal P}_{\cal R}\simeq \frac{V^3}{12\pi^2 V_{,\phi}^2}{\cal A}\simeq\frac{V^3}{12\pi^2 V_{,\phi}^2}\Big(1+\theta(\phi)V\Big).
\end{equation}
The observational restriction on  amplitude of the scalar power spectrum is quantified by  Planck
collaboration from the anisotropies of cosmic microwave background (CMB) at pivot scale $(k_{*}=0.05~\rm Mpc^{-1})$ \citep{akrami:2018} as
\begin{equation}\label{psrestriction}
  {\cal P}_{\cal R}(k_{*})\simeq 2.1 \times 10^{-9}.
\end{equation}
The association of scalar spectral index $n_{s}$ with the slow-roll parameters in the NMDC model can be computed from the curvature power spectrum through the definition $n_{s}-1\equiv d\ln{\cal P}_{\cal R}/d\ln k$   \cite{Teimoori:2021} as follows
\begin{align}\label{nsSR}
n_s\simeq  1-\frac{1}{{\cal A}}\left[6\varepsilon_{V}-2\eta_{V}+2\varepsilon_{V}\left(1-\frac{1}{{\cal A}}\right)\right.
 \left.
\left(1+\frac{\theta_{,\phi}}{\theta(\phi)}\frac{V(\phi)}{V_{,\phi}}\right)\right],
\end{align}
in which
\begin{equation}\label{eta}
\eta_{V}=\frac{V_{,\phi\phi}}{V}.
\end{equation}
In the case of  $\theta(\phi)=0$ the coupling parameter is faded away  and  standard slow-roll inflationary formalism is retrieved.
The tensor power spectrum at $c_{t}k=aH$ and the
tensor-to-scalar ratio in NMDC framework under slow-roll approximation have been computed in \cite{Tsujikawa:2013} as the following form
\begin{align}\label{PtSR}
&{\cal P}_{t}=\frac{2H^2}{\pi ^{2}},\\
\label{r}
&r\simeq 16 \varepsilon \simeq 16 \frac{\varepsilon_V}{{\cal A}}.
\end{align}
The observational constraint on the scalar spectral index $n_s$ in accordance with Planck  2018  TT,TE,EE+lowE+lensing+BK14
+BAO data at the 68\%  CL , and the upper limit on the tensor-to-scalar ratio $r$ at 95\% CL \cite{akrami:2018} are as follows
\begin{align}\label{nsconsraint}
&n_s= 0.9670 \pm 0.0037,\\
\label{rconsraint}
&r<0.065.
\end{align}
%====================================Intensification of the Curvature Perturbation========================
\section{Amplification of  Curvature  Power Spectrum}\label{sec3}
It is corroborated that, generation of observable PBHs and GWs originates from remarkable enhancement  in the amplitude of curvature power spectrum during transient USR phase on small scales.  In NMDC framework the appropriate model parameters and coupling function between field derivative and the Einstein tensor should be elected so as to enhance the friction during USR era  \cite{fu:2019,Teimoori:2021,fu:2020,Dalianis:2020,Heydari:2021}. In pursuance of this objective we delineate two-parted exponential form of coupling function $\theta(\phi)$ so as to have a NMDC model with  prognostications in conformity with the latest observational data on large scales (CMB) as well a generated peak in the scalar power spectrum on  smaller  scales, as follows
\begin{equation}\label{t}
\theta(\phi)=\theta_I(\phi)\Big(1+\theta_{II}(\phi)\Big),
\end{equation}
in which
\begin{align}
\label{tI}
&\theta_I(\phi)=\frac{e^{\alpha\phi}}{M^{2}},\\
\label{tII}
&\theta_{II}(\phi)=\frac{\omega}{\sqrt{\left(\frac{\phi-\phi_c}{\sigma}\right)^2+1}}\,.
\end{align}
The first portion of our coupling function (\ref{tI}) is a generic exponential form of the taken coupling  parameters by \cite{Defelice:2011,Granda:2020,Tsujikawa:2012,Tsujikawa:2013}. Another portion (\ref{tII}) is given by Fu et al. \cite{fu:2019} so as to inspect the possibility of PBHs and GWs generation in a transient NMDC framework for the potential $V(\phi)\propto\phi^{2/5}$, however we combine these two functions by way of (\ref{t}). Apropos of $\theta_{II}(\phi)$, this function has an acme at crucial value of field  $\phi=\phi_{c}$ with the height and width denoted  by $\omega$ and $\sigma$. It can be inferred from (\ref{tII}) that, for  farther field values from the crucial value $\phi_{c}$ the function $\theta_{II}(\phi)$  melts away and our general coupling function (\ref{t}) is dominated by its first term (\ref{tI}). The presentment of exponential function $\theta_{I}(\phi)$ is necessitated in pursuance of rectifying  prognostications of our model on the CMB scales with the recent observational data. Furthermore,  $\theta_{I}(\phi)\theta_{II}(\phi)$ beside  fine tuning  of model parameters are productive of increase in the amplitude of curvature power spectrum on small scales to sufficient value to produce the observable PBHs and GWs. Concerning the portions  of coupling function (\ref{tI})-(\ref{tII}), parameters  $\{\alpha, \omega\}$ are dimensionless whereas the parameters $\{\phi_{c},\sigma, M\}$ have  dimensions of mass.
Inasmuch as, in standard framework of inflation the exponential potential drives an endless inflationary epoch \cite{karami:2017} and, as regards the inconsistency of the prognostications of this form of potential at CMB scales  in view of  Planck  2018  TT,TE,EE+lowE+lensing+BK14+BAO data \cite{akrami:2018,karami:2017}, we try to amend the results of this potential in NMDC framework.
Ergo, in the following we pursue the objective if the exponential potential  could  derive the viable inflation on large scale  contemporaneous with production detectable PBHs and GWs on smaller scales  in NMDC framework. The exponential potential is given as follows
\begin{equation}\label{v}
V(\phi)=\lambda e^{k\phi},
\end{equation}
where $k$ is dimensionless parameter, and  $\lambda$ can be fixed through the constraint of scalar  power spectrum at pivot scale $k_{*}$ (\ref{psrestriction})  in our setup.

In this stage we ponder over the  estimation of the order of required multiplication factor to increase the amplitude of the scalar power spectrum ${\cal P}_{\cal R}$ on small scales to order ${\cal O}(10^{-2})$
\begin{table}
  \centering
  \caption{Tuned parameters for the Cases A, B, C, and D. Also $\Delta N$ denotes the duration of viable inflationary epoch for each Case. The parameter $\lambda$ is specified by the power spectrum constraint (\ref{psrestriction}) at horizon traversing  $e$-fold number $(N_{*})$.}
\begin{tabular}{cccccc}
  \hline
 $\#$ &\qquad $\omega$\qquad &\qquad$\sigma$\qquad & \qquad$\phi_{c}$\qquad&\qquad $\lambda$\qquad&\qquad $\Delta N$\qquad\\[0.5ex] \hline\hline
  Case A& \qquad$3.869\times10^{7}$\qquad &\qquad$1.53\times10^{-11}$\qquad &\qquad$0.1272$ \qquad& \qquad $3.68\times10^{-10}$\qquad&\qquad$53$ \qquad \\[0.5ex] \hline
  Case B&\qquad $3.879\times10^{7}$\qquad &\qquad$1.67\times10^{-11}$\qquad &\qquad$0.125$\qquad &\qquad$3.71\times10^{-10}$ \qquad&\qquad$58$ \qquad \\ \hline
  Case C&\qquad$5.146\times10^{7}$\qquad &\qquad $1.86\times10^{-11}$\qquad & \qquad$0.117$\qquad& \qquad$3.68\times10^{-10}$ \qquad&\qquad$60$ \qquad\\ \hline
  Case D&\qquad$5.761\times10^{7}$\qquad &\qquad $2.17\times10^{-11}$\qquad &\qquad $0.112$ \qquad&\qquad $3.63\times10^{-10}$\qquad&\qquad$60$ \qquad \\ \hline
\end{tabular}
 \label{tab1}
\end{table}
\begin{table}[H]
  \centering
  \caption{Computed numerical upshots  for Cases of Table \ref{tab1}. The quantities ${\cal P}_{ \cal R}^\text{peak}$,  $f_{\text{PBH}}^{\text{peak}}$, and $M_{\text{PBH}}^{\text{peak}}$ denote  the  values of the scalar  power spectrum,  fractional abundance of  PBHs, and  the PBH mass at acme position $\phi=\phi_{c}$, respectively. The numerical results of  $n_{s}$ and $r$ are calculated  at horizon traversing  CMB $e$-fold number $(N_{*})$.}
\begin{tabular}{ccccccc}
  \hline
   $\#$ & \qquad $n_{s}$\qquad &\qquad $r$\qquad &\qquad ${\cal P}_{ \cal R}^\text{peak}$\qquad &\qquad$k_{\text{peak}}/\text{Mpc}^{-1}$\qquad& \qquad$f_{\text{PBH}}^{\text{peak}}$\qquad& \qquad$M_{\text{PBH}}^{\text{peak}}/M_{\odot}\qquad$\\ \hline\hline
  Case A &\qquad0.9731\qquad  &\qquad0.0419\qquad& \qquad0.050\qquad & \qquad$3.52\times10^{6}$ \qquad&\qquad0.0012\qquad &\qquad$19.10$\qquad \\ \hline
 Case B &\qquad 0.9701\qquad  & \qquad0.0373\qquad &\qquad0.0423 \qquad&\qquad  $5.70\times10^{8}$ \qquad&\qquad0.0355\qquad &\qquad $7.28\times10^{-6}$ \qquad\\ \hline
 Case C &\qquad0.9696\qquad  & \qquad0.0360\qquad & \qquad0.034\qquad & \qquad$2.95\times10^{12}$\qquad &\qquad 0.9615\qquad &\qquad$2.713\times10^{-13}$ \qquad\\ \hline
 Case D &\qquad 0.9689\qquad  &\qquad0.0367\qquad &\qquad0.0312\qquad & \qquad$4.812\times10^{13}$ \qquad& \qquad0.9526 \qquad&\qquad$1.023\times10^{-15}$\qquad \\ \hline
\end{tabular}
\label{tab2}
\end{table}
\noindent
 in comparison with CMB scales,  which is sufficient value to generate PBHs. Thus, in the following  an estimated connection between  ${\cal P}_{\cal R}$ at $\phi=\phi_{c}$ and  ${\cal P}_{\cal R}$ at $\phi=\phi_{*}$ is inferred, as regards  the field value at the moment  of horizon traversing by pivot scale denotes by $\phi_{*}$.  Utilizing  equations (\ref{PsSR}) and (\ref{t})-(\ref{v}) we obtain
 \begin{equation}\label{PsSRHiggs}
   {\cal P}_{\cal R}\simeq \frac{\lambda e^{k\phi}}{12k^{2} \pi^{2}}\left[1+ q e^{(k+\alpha)\phi} \left(1+\frac{\omega}{\sqrt{1+(\frac{\phi-\phi_{c}}{\sigma})^{2}}}\right)\right],
 \end{equation}
 where
 \begin{equation}\label{q}
   q\equiv\frac{\lambda}{M^{2}},
 \end{equation}
 supposing the following proviso
 \begin{equation}\label{assumptions}
\omega\gg 1, \hspace{1.5cm} \mid \phi_{*}-\phi_{c}\mid\, \gg \sigma \omega ,
\end{equation}
after all, in a rough approximation we conclude
\begin{equation}\label{Pspeak}
{\cal P}_{\cal R}\Big|_{\phi= \phi_{c}}\simeq \omega\times{\cal P}_{\cal R}\Big|_{\phi=\phi_{*}}.
\end{equation}
As regards  ${\cal P}_{\cal R}$ at the moment of horizon crossing (\ref{psrestriction}), for $\omega\sim{\cal O}(10^{7})$ the acme  of scalar  power spectrum at $\phi=\phi_{c}$ could increase to order ${\cal O}(10^{-2})$. Taking the above approximation (\ref{Pspeak}) and the mentioned provisoes  (\ref{assumptions}) into account,  we specify  $\alpha=40$, $q=20$, $k=8$, and  adjust four disparate  parameter Cases listed in Table \ref{tab1}.  It should be noting that, our model is delineated by a collection of eight   parameters as   $\{ k, \alpha, M,   \lambda,  q, \omega, \phi_c, \sigma\}$, whereas  the parameters  $M$, $q$, and $\lambda$ are associated together through equation (\ref{q}).   Table \ref{tab2} embodies  the  computed  numerical results for quantities pertinent to  inflation  $n_{s}$, $r$, and the ones  affiliated to  PBHs  formation.

\begin{figure*}
\begin{minipage}[b]{1\textwidth}
\vspace{-1.cm}
\subfigure{\includegraphics[width=.48\textwidth]%
{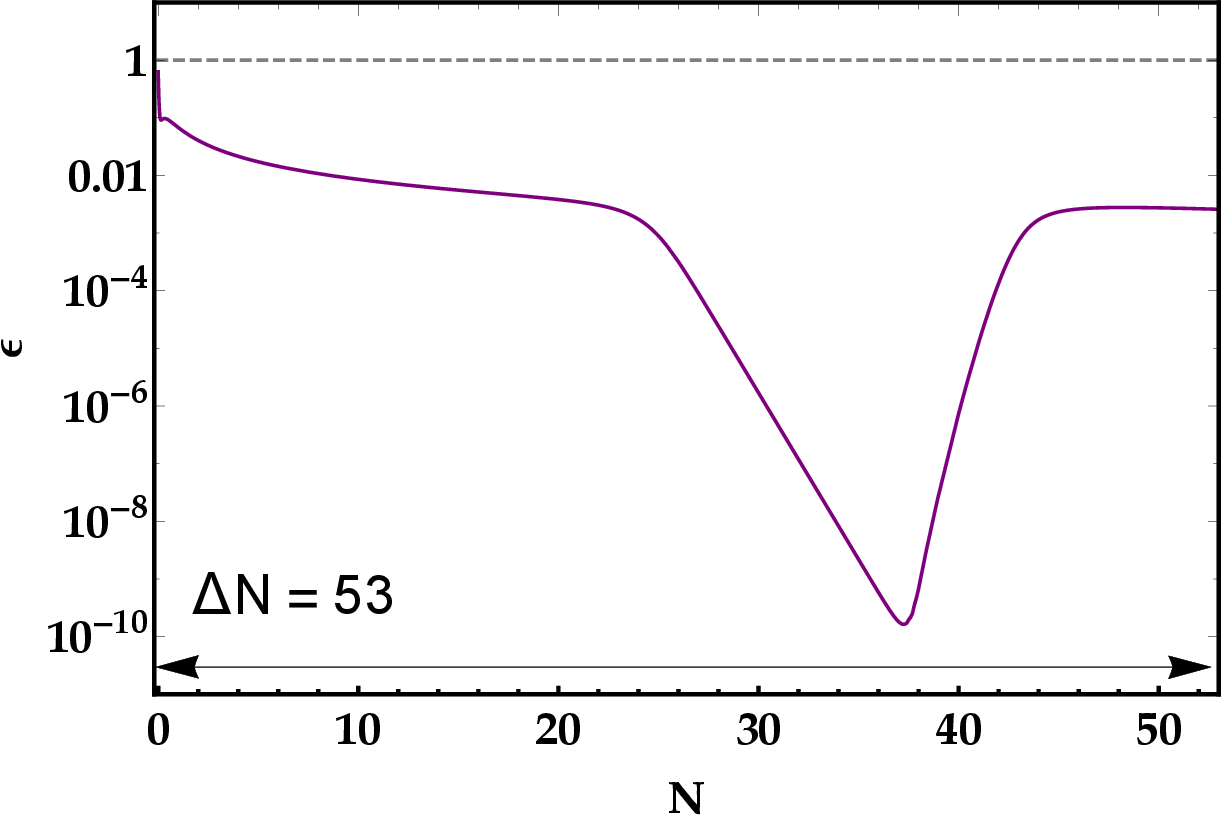}} \hspace{.1cm}
\subfigure{ \includegraphics[width=.48\textwidth]%
{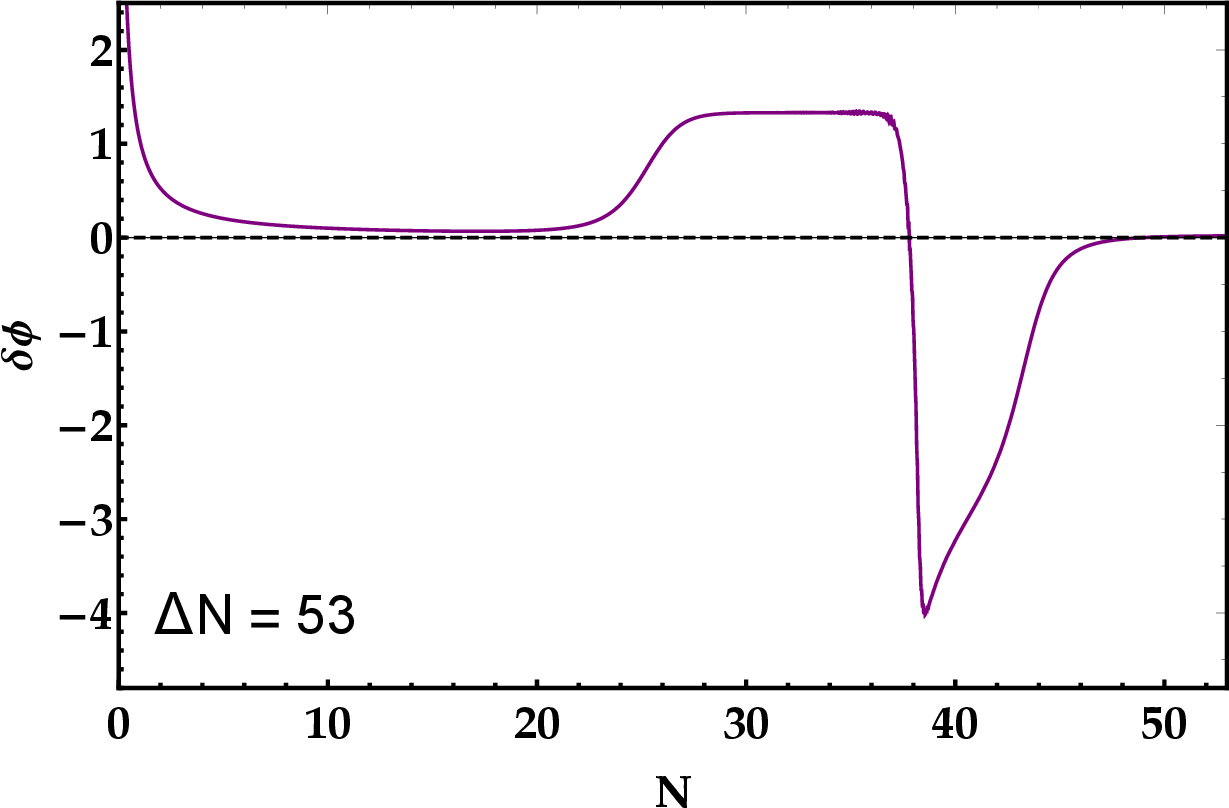}}\hspace{.1cm}
\subfigure{ \includegraphics[width=.48\textwidth]%
{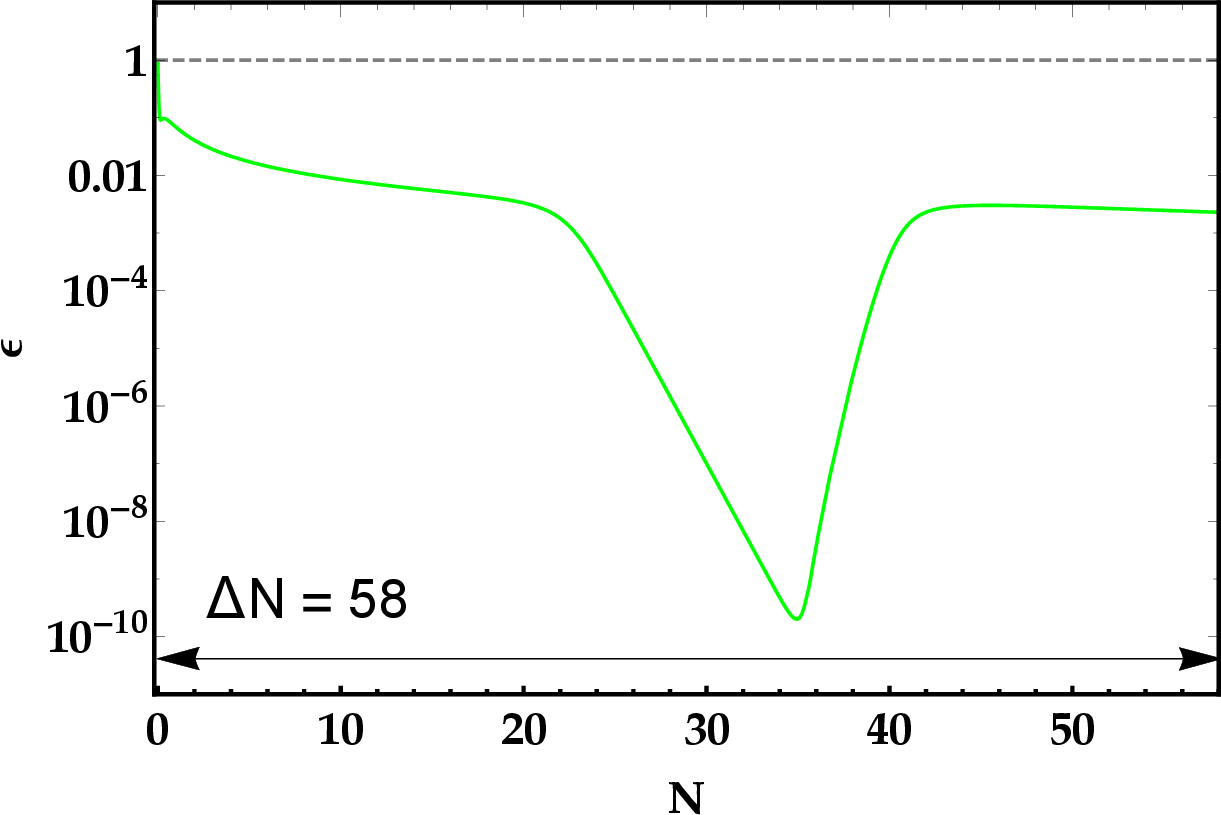}}\hspace{.1cm}
\subfigure{ \includegraphics[width=.48\textwidth]%
{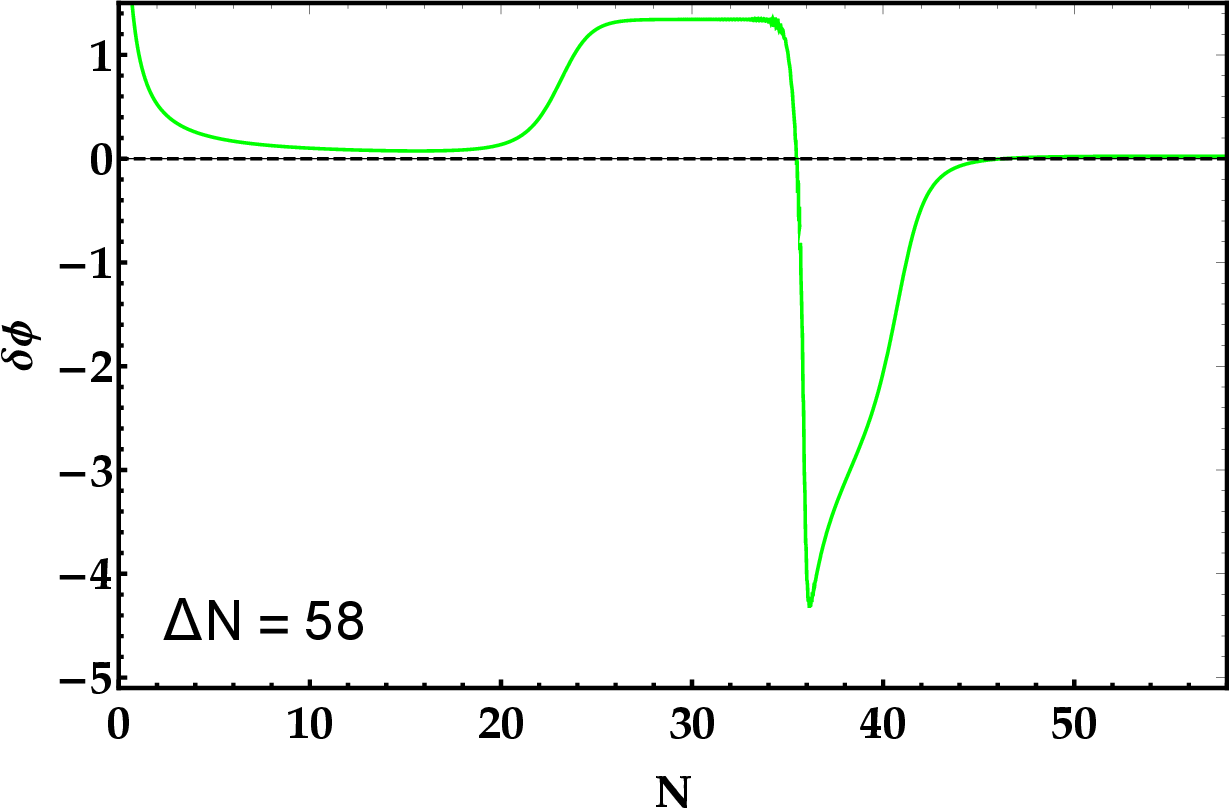}} \hspace{.1cm}
\subfigure{\includegraphics[width=.48\textwidth]%
{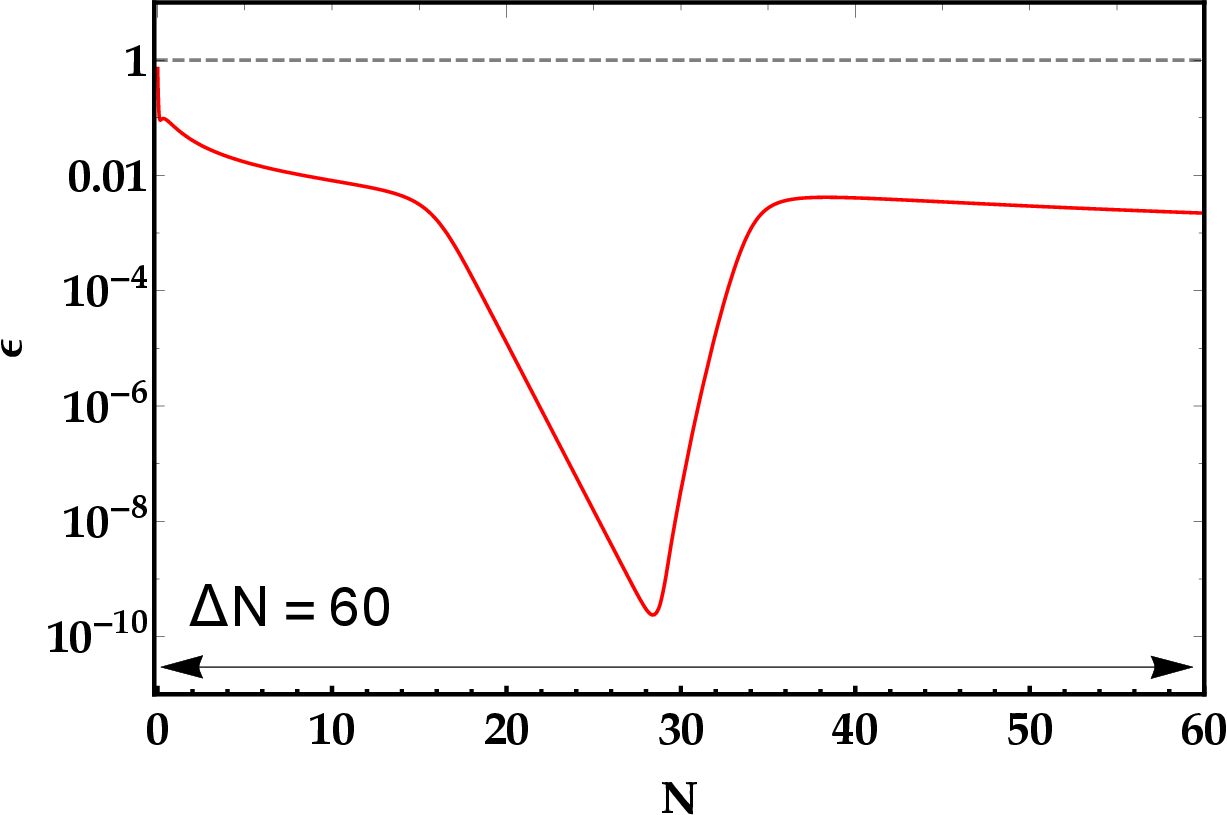}} \hspace{.1cm}
\subfigure{ \includegraphics[width=.48\textwidth]%
{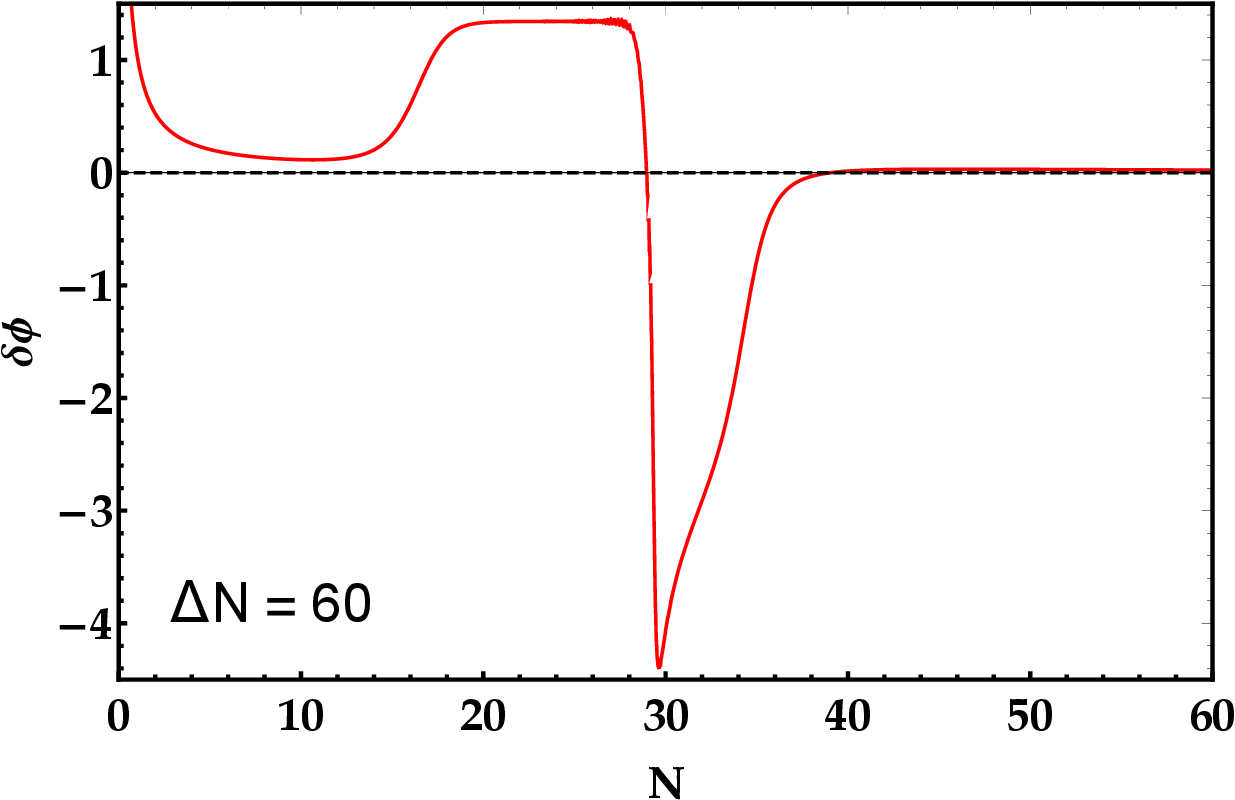}}\hspace{.1cm}
\subfigure{ \includegraphics[width=.48\textwidth]%
{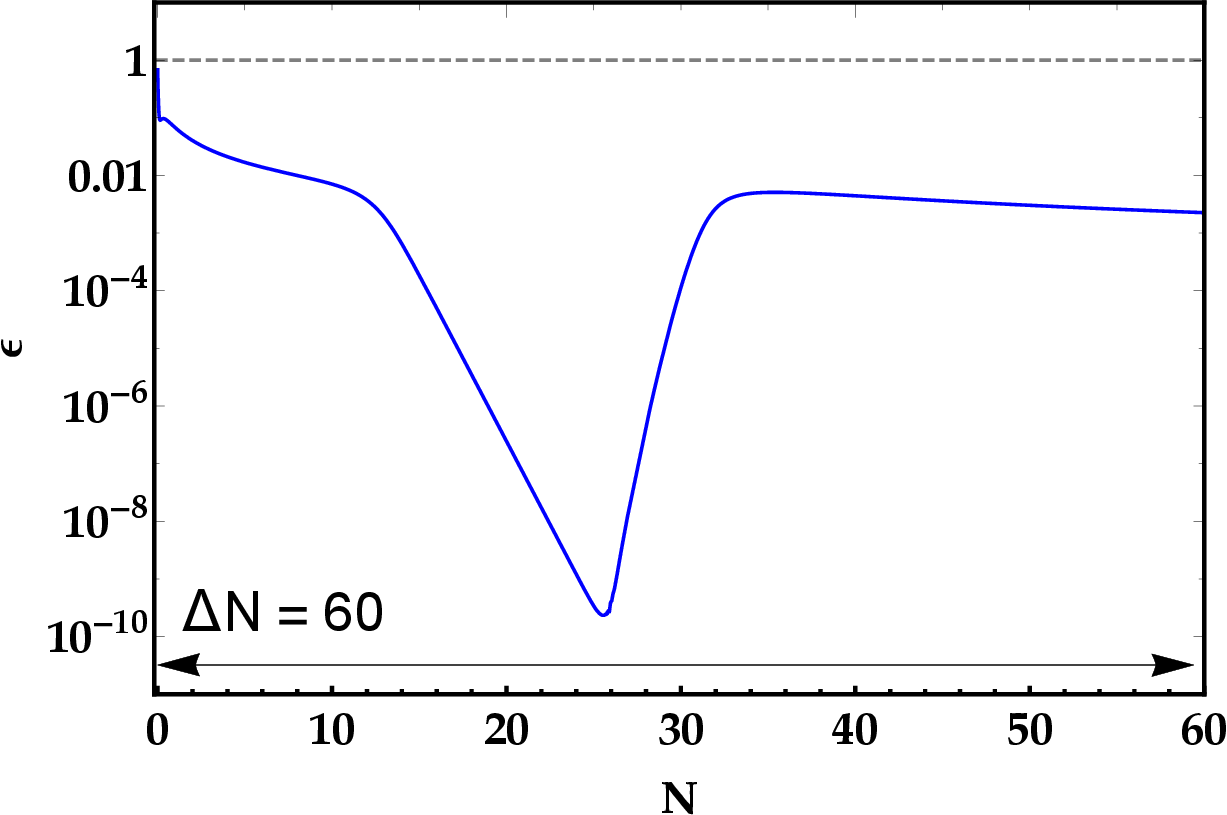}}\hspace{.1cm}
\subfigure{ \includegraphics[width=.48\textwidth]%
{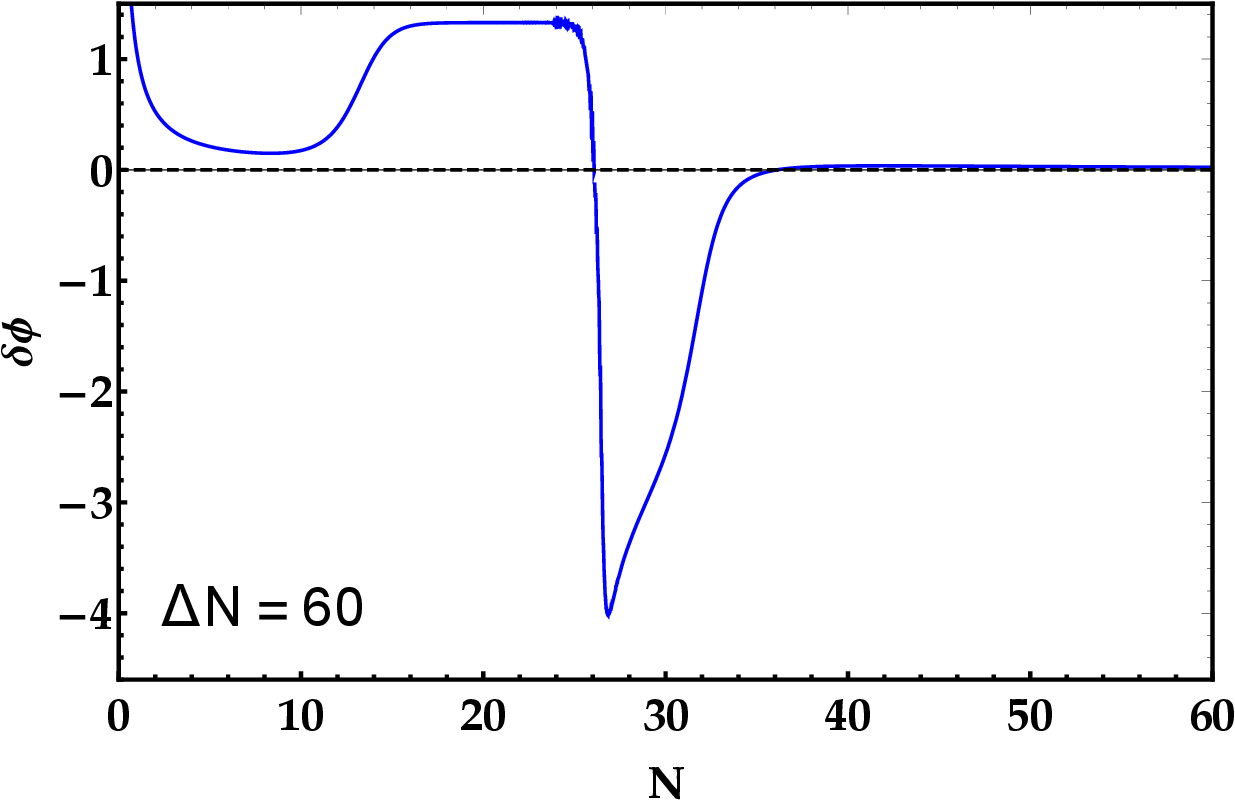}}
\end{minipage}
\vspace{-1.cm}
\caption{ Variation of (left) the first slow-roll parameter $\varepsilon$, (right) the second slow-roll parameter $\delta_{\phi}$, with regard to the $e$-fold number $N$ for the Cases A, B, C, and D represented by purple,  green,  red, and blue lines, respectively.}
\label{fig:SRp}
\end{figure*}

\begin{figure}[H]
\centering
\vspace{-0.2cm}
\scalebox{0.6}[0.6]{\includegraphics{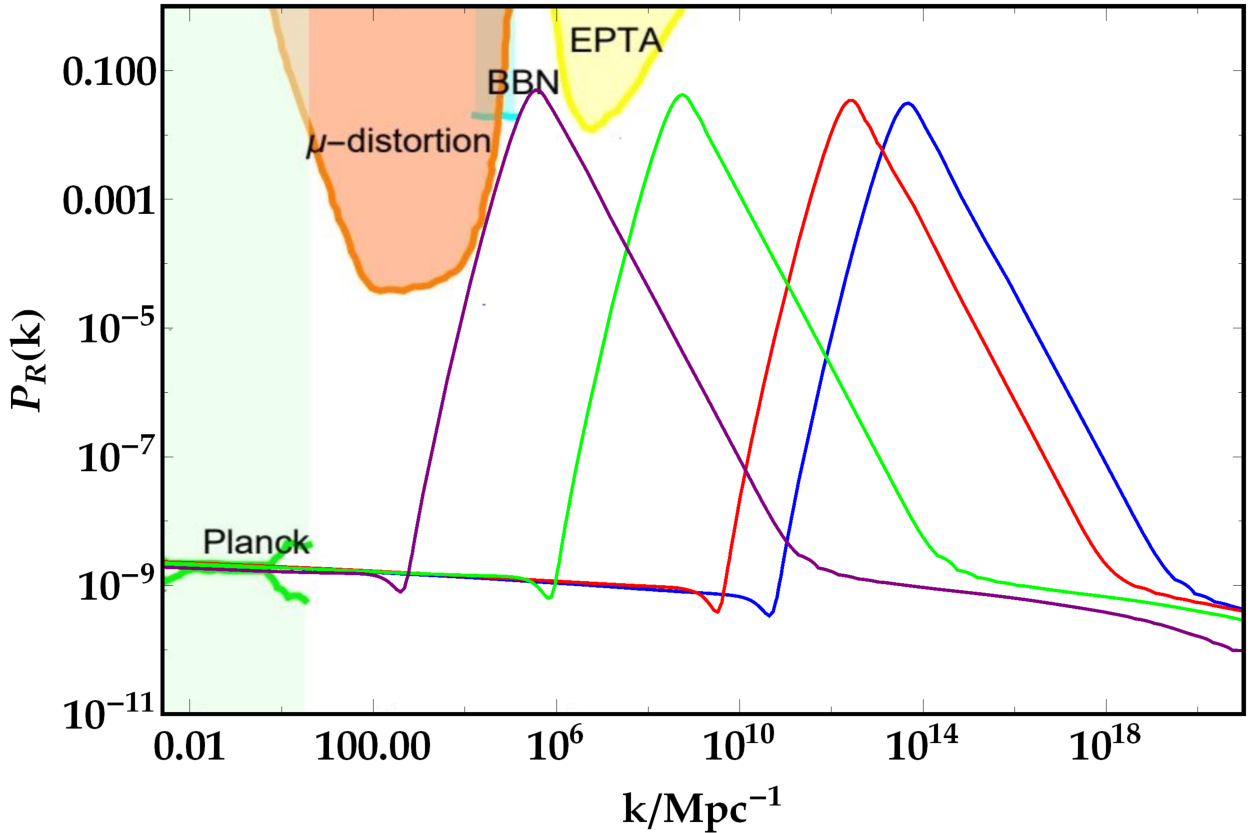}}
\vspace{-0.6cm}
\caption{The attained precise curvature fluctuations power spectrum with regard to  comowing  wavenumber $k$ ensued from  resolving the Mukhanov-Sasaki equation numerically. The purple, green, red and blue lines appertain to the Cases A, B, C, and D, respectively.  The light-green, yellow, cyan, and orange  shadowy domains  depict the confinements of  CMB observations \cite{akrami:2018},  PTA observations \cite{Inomata:2019-a},   the effect on the ratio between neutron and proton during the big bang nucleosynthesis (BBN) \cite{Inomata:2016}, and the $\mu$-distortion of CMB \cite{Fixsen:1996}, respectively.}
\label{fig-ps}
\end{figure}

As regards the duration of observable inflationary epoch thereabout 50-60 $e$-folds number from the time of horizon traversing via pivot scale $k_{*}$ to the end of inflation, and so as to have a viable inflationary era, we adjust the $e$-fold number of  horizon traversing  $N_{*}$ as $53$ for Case A, $58$ for Case B, and $60$  for Cases C and D. The schemed results in Fig. \ref{fig:SRp} for the first slow-roll parameter illustrate that, the end of inflationary era is stipulated at the $e$-fold number $N_{\text{end}}=0$ for all Cases of our model through resolving $\varepsilon =1$.
As we mentioned in the preceding section, the first slow-roll parameter $\varepsilon$ in our NMDC setup can be approximated by equation (\ref{epsilon}), in which the presence of coupling function $\theta(\phi)$ by way of ${\cal A}$ leads to bring the $\varepsilon$ to one at $N_{\text{end}}=0$ and terminate the inflationary era for exponential potential (\ref{v}). It is obvious from equation (\ref{epsilon}) that, for $\theta(\phi)=0$ and ${\cal A}=1$ the standard slow-roll inflationary model is retrieved with $\varepsilon=\varepsilon_{v}=k^{2}/2$, which is a constant value and leads to endless inflation driven by exponential potential.
\begin{figure*}
\begin{minipage}[b]{1\textwidth}
\vspace{-1.cm}
\subfigure{\includegraphics[width=.48\textwidth]%
{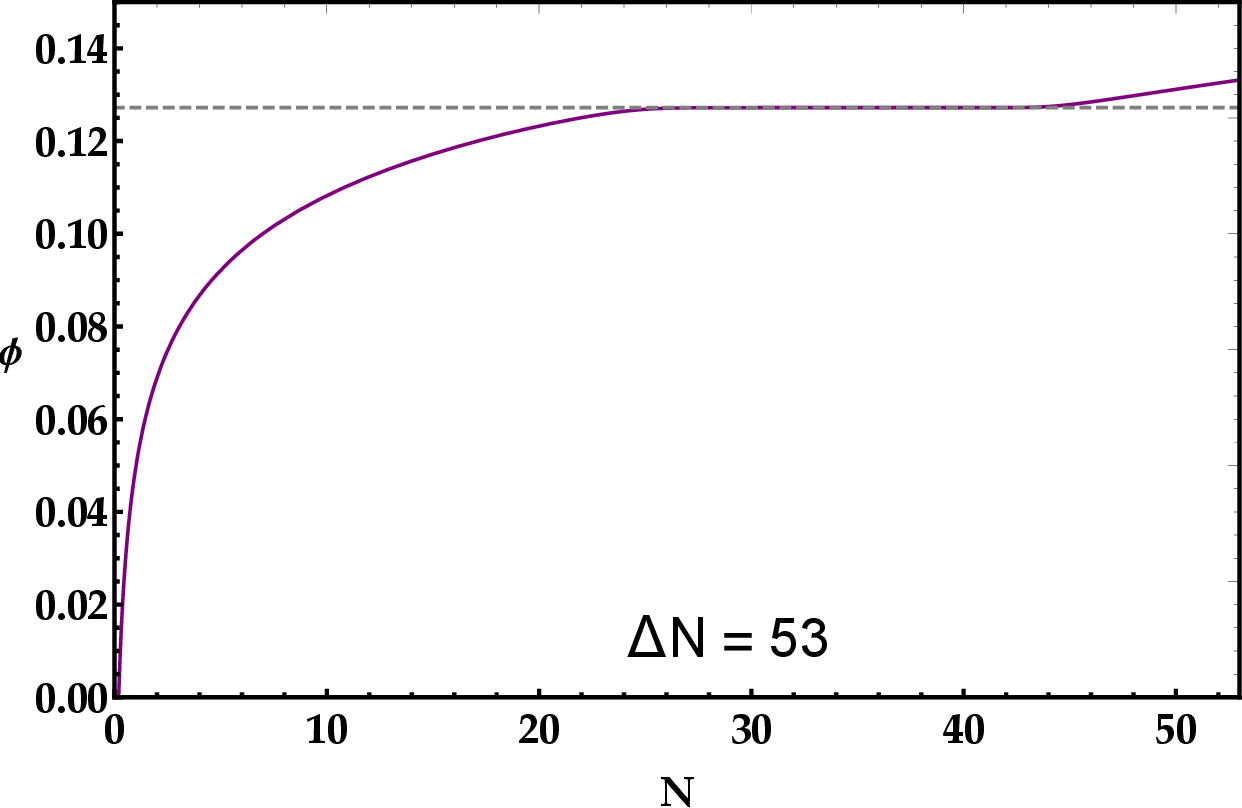}} \hspace{.1cm}
\subfigure{ \includegraphics[width=.48\textwidth]%
{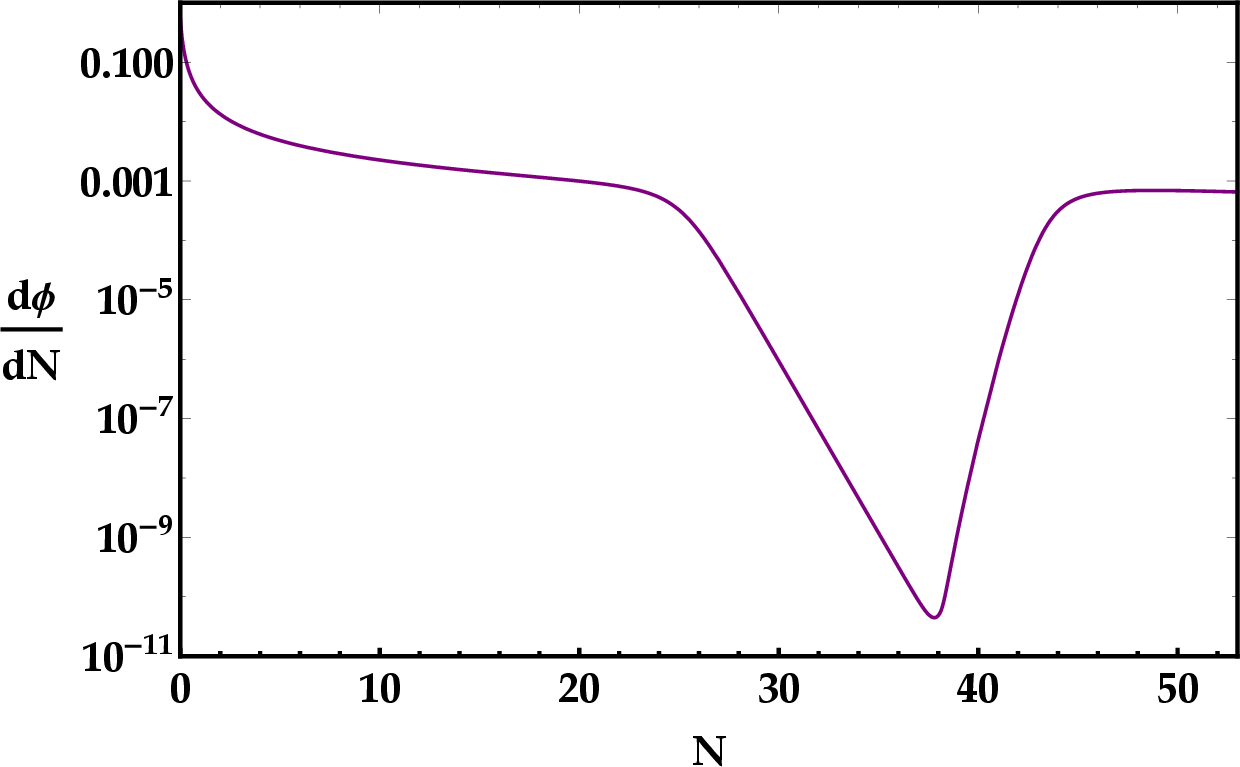}}\hspace{.1cm}
\subfigure{ \includegraphics[width=.48\textwidth]%
{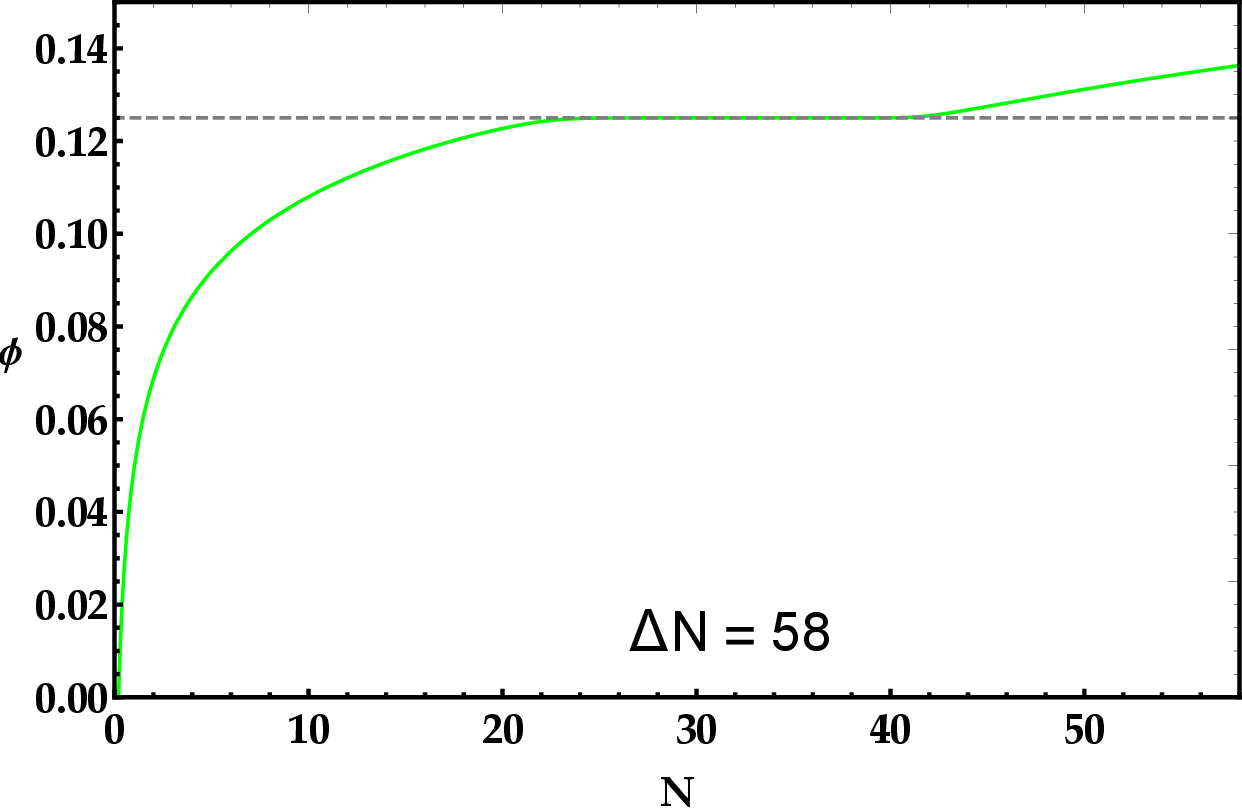}}\hspace{.1cm}
\subfigure{ \includegraphics[width=.48\textwidth]%
{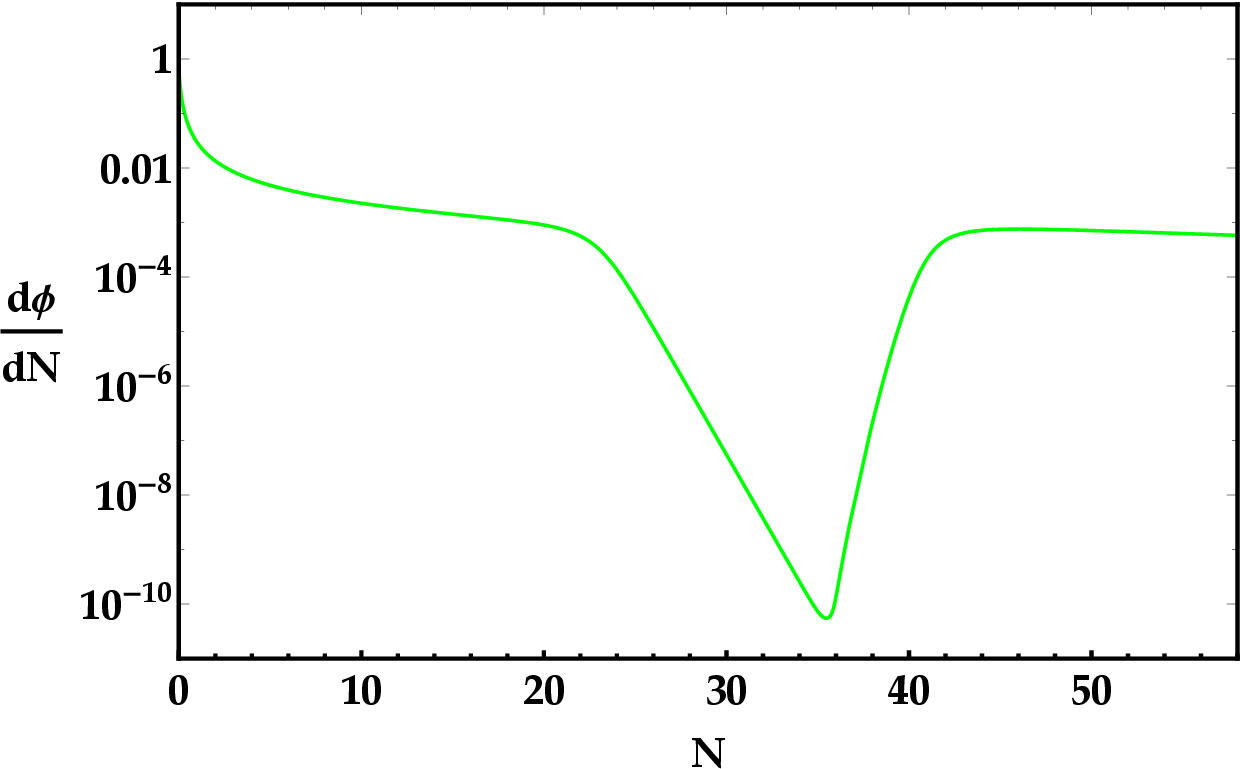}} \hspace{.1cm}
\subfigure{\includegraphics[width=.48\textwidth]%
{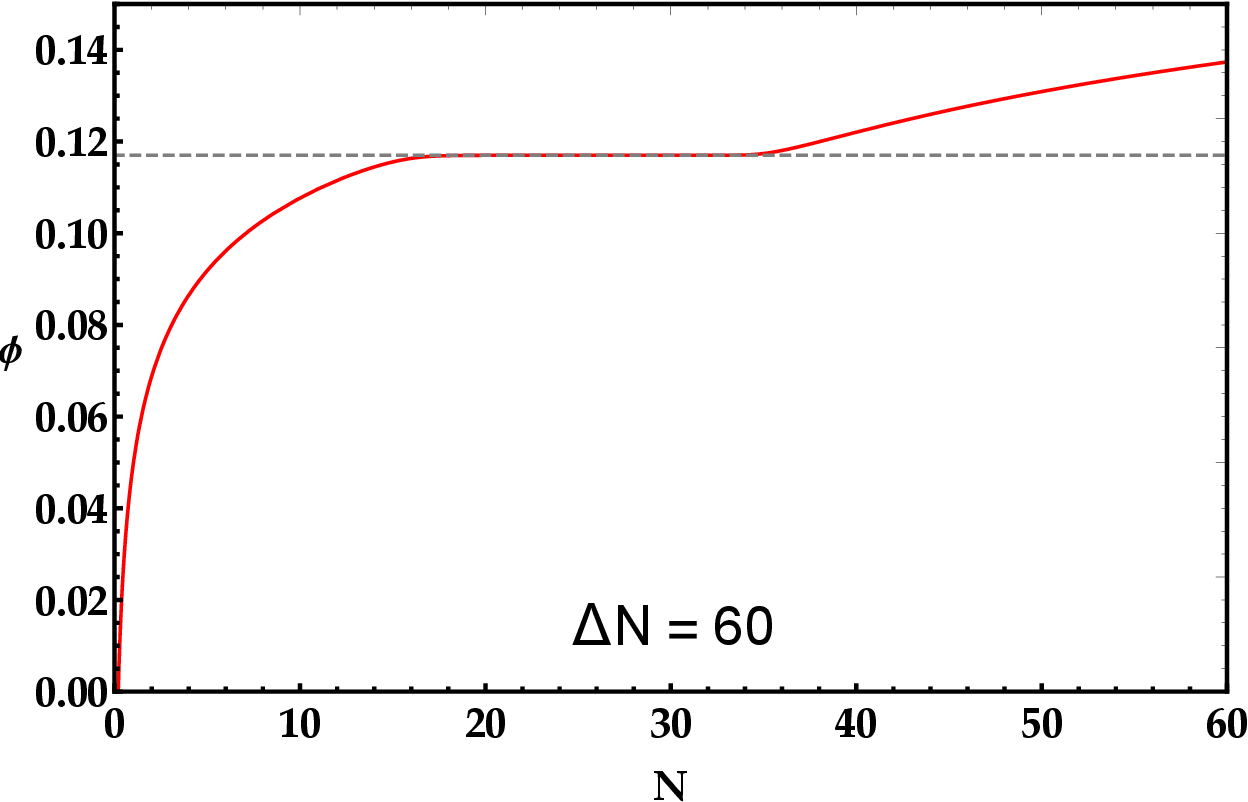}} \hspace{.1cm}
\subfigure{ \includegraphics[width=.48\textwidth]%
{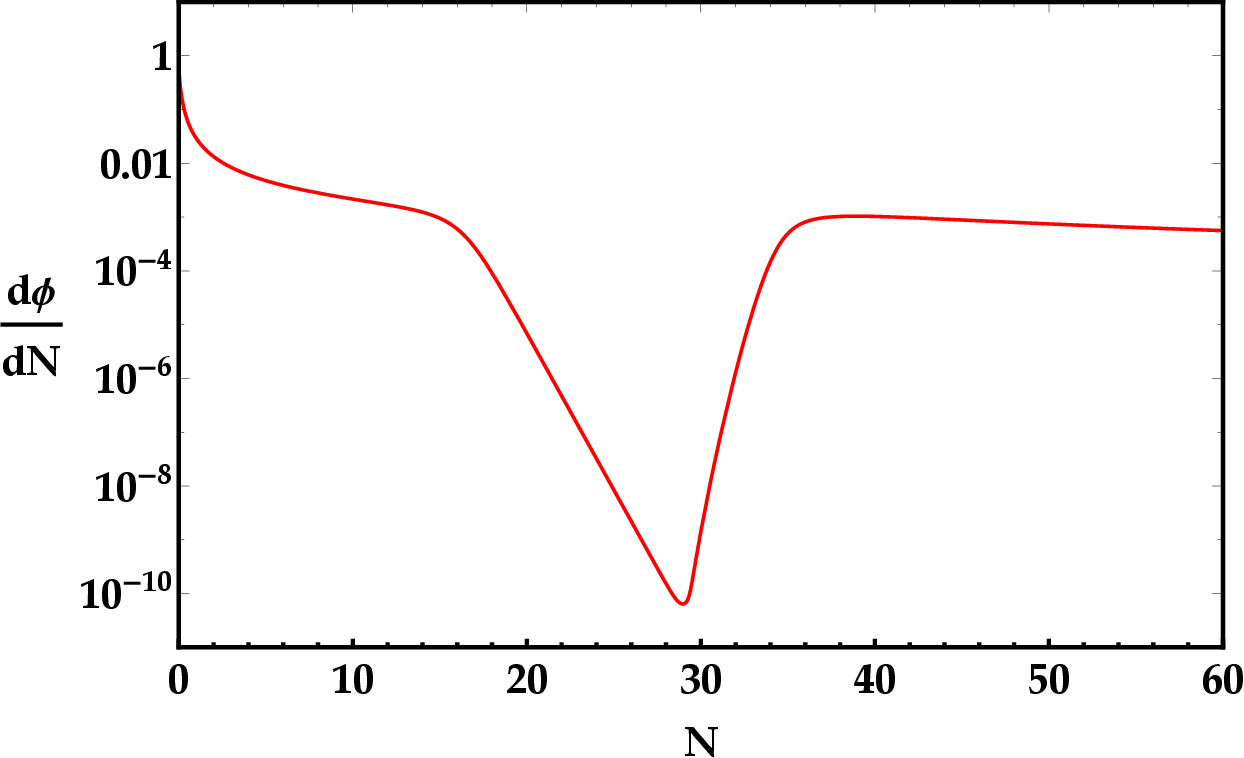}}\hspace{.1cm}
\subfigure{ \includegraphics[width=.48\textwidth]%
{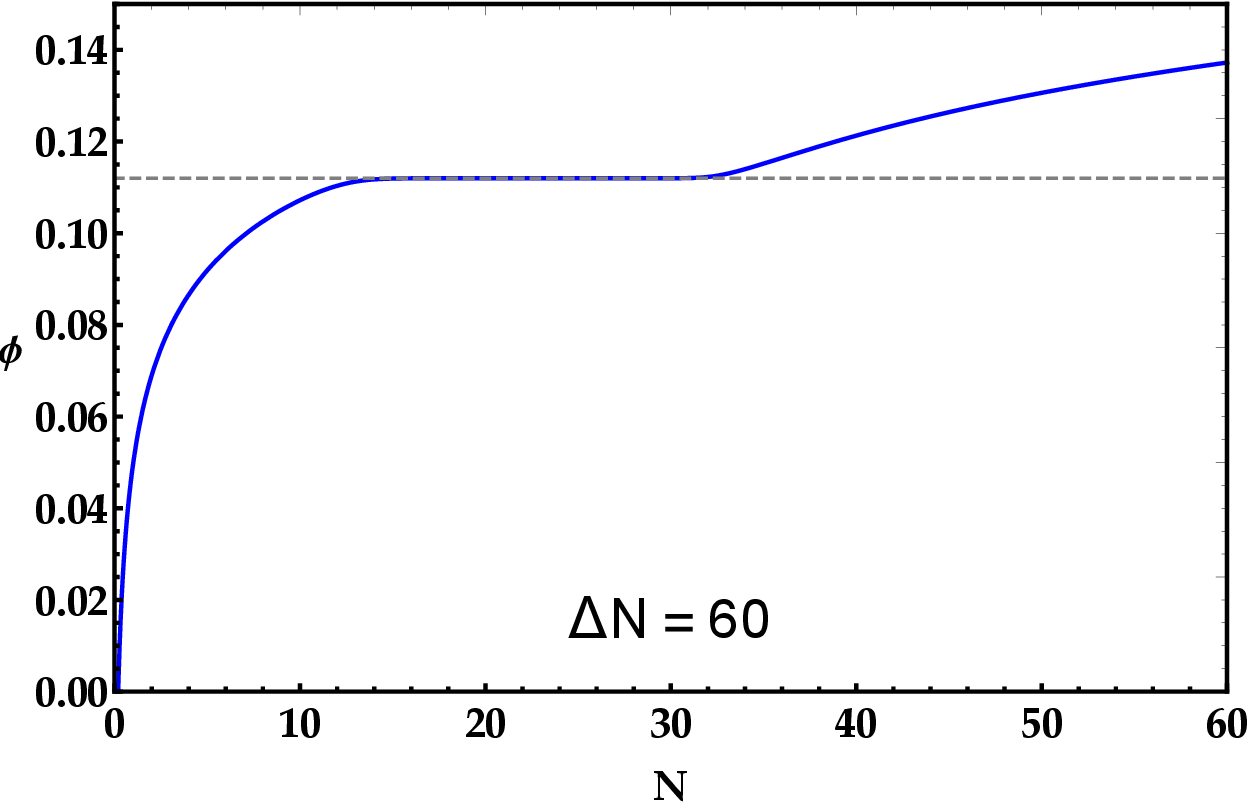}}\hspace{.1cm}
\subfigure{ \includegraphics[width=.48\textwidth]%
{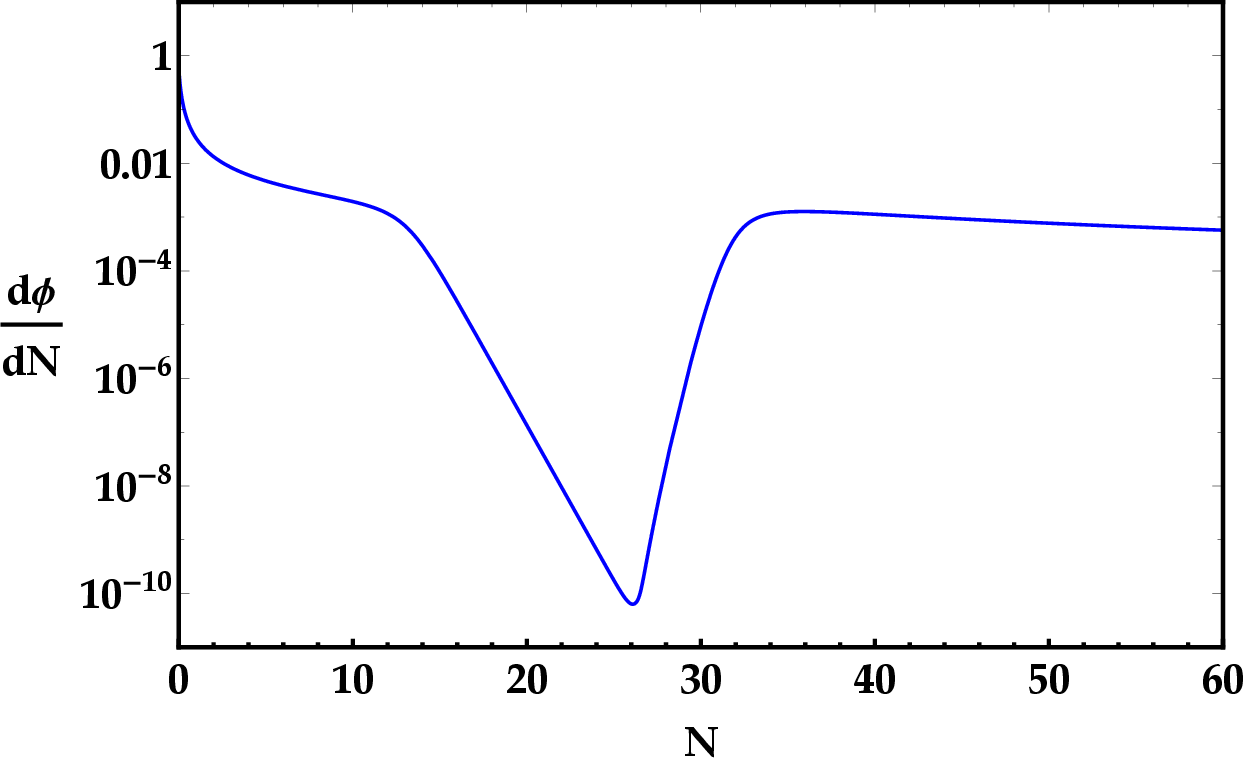}}
\end{minipage}
\vspace{-1.cm}
\caption{(left) Variation of the scalar field $\phi$, (right) the velocity of the scalar field $\frac{d\phi}{dN} $  against the $e$-fold number $N$ for the Cases A, B, C, and D represented by purple,  green,  red, and blue lines, respectively. The acme position $\phi=\phi_{c}$ is represented by dashed line for all Cases in left plots, and  slow-roll approximation Eqs. (\ref{FR1:SRC}) and (\ref{Field:SRC}) are utilized for the incipient conditions.}
\label{fig:phi}
\end{figure*}
In pursuance of comprehending the evolution of inflaton field  versus $e$-folds number we need to unravel  the background equations (\ref{FR1:eq})-(\ref{Field:eq}) exactly, utilizing the exponential potential (\ref{v}) and  NMDC coupling function (\ref{t})-(\ref{tII}). Thereafter, in Fig. \ref{fig:phi} we exhibit the accurate status of the inflaton  field  $\phi$ and its velocity $\frac{d\phi}{dN}$ with regard to  $e$-folds number $N$ ($dN=-Hdt$)  from $N_{*}$ the horizon traversing  to $N_{\text{end}}$ the end of inflation,  for the Cases A, B, C, and D  by purple, green, red, and blue lines. For all Cases of this figure, There is  an ephemeral  smooth domain  (Ultra Slow-Roll phase) continuing  for about  20 $e$-folds  in the vicinity of  $\phi=\phi_{c}$. By reason of dominance of intensified friction during this  USR phase the inflaton field rotates in a very leisurely way and the slow-roll proviso is violated, thus the sufficient time to enhance the amplitude of curvature power spectrum to around order ${\cal O}(10^{-2})$ can be provided.
 The sever decline in the  velocity of the inflaton field during the USR phase is obvious from the right plots of  Fig. \ref{fig:phi} for all Cases.
Hereupon, intensification of curvature power spectrum at small scale in the course of USR period is schemed in Fig. \ref{fig-ps}.

The mutation of slow-roll parameters $\varepsilon$ and $\delta_{\phi}$ with regard to the $e$-fold number $N$ in the course of observable inflationary epoch $\Delta N=N_{*}-N_{\text{end}}$ are represented in left and right plots of Fig. \ref{fig:SRp} for all Cases of Table \ref{tab1}. With regard to left plots of Fig. \ref{fig:SRp}, the severe diminution in the value of first slow-roll parameter  $\varepsilon$ to around order ${\cal O}(10^{-10})$ in  USR domain occurs for all Cases of our model, that can lead to intensification  in the amplitude of  scalar  power spectrum. Furthermore, from left plots of Fig. \ref{fig:SRp}, it can be inferred that the mentioned slow-roll proviso is contravened  via the second slow-roll parameter  $\delta_{\phi}$ through exceeding one in the course of USR stage momentarily.
Note that it can be seen from  the mutation of  slow-roll parameters in  Fig. \ref{fig:SRp} that, the slow-roll provisos at horizon traversing  $e$-fold number $N_{*}$ are satisfied for each Case,  thereupon Eqs. (\ref{nsSR}) and (\ref{r}) can be applied to compute the scalar spectral index $n_{s}$ and tensor-to-scalar ratio $r$ in our framework. The numerical computations epitomized in Table \ref{tab2} corroborate that in view of Planck  2018  TT,TE,EE+lowE+lensing+BK14+BAO data the quantity of
$r$ for the entire Cases and $n_{s}$ for the Cases B, C, and D are consistent with $68\%$ CL, whereas $n_{s }$ for Case A is  consonant   with $95\%$  CL \cite{akrami:2018}. As a consequence of choosing NMDC groundwork and appropriate exponential form of coupling parameter for our model, we could rectify the observational prognostications of exponential potential.

By reason of invalidity of slow-roll approximation in USR domain, we cannot utilize Eq.
 (\ref{PsSRHiggs}) to compute the curvature fluctuations power spectrum in this stage owing to deriving from slow-roll approximation. Thereupon so as to attain the precise value of curvature power spectrum, the numerical evaluation of the succeeding  Mukhanonv-Sasaki (MS) equation is necessitated  for the entire Fourier modes
\begin{equation}\label{MS}
 \upsilon^{\prime\prime}_{k}+\left(c_{s}^2 k^2-\frac{z^{\prime\prime}}{z}\right)\upsilon_k=0,
\end{equation}
in which the prime denotes  derivative with regard to the conformal time $\eta=\int {a^{-1}dt}$, and
\begin{equation}\label{z}
 \upsilon\equiv z {\cal R}, \hspace{1cm} z=a\sqrt{2Q_s}.
\end{equation}
It is worth noting that, the mutation of curvature fluctuations ${\cal R}$   in Fourier space $\upsilon_k$ in the course of inflationary epoch from sub-horizon scales $c_{s}k\gg aH$ to  super-horizon scales  $c_{s} k\ll aH$ is evaluated by MS equation (\ref{MS}).
We appoint the  Fourier transformation of  the  Bunch-Davies vacuum state as the incipient proviso at the sub-horizon scale \cite{Defelice:2013} as follows
\begin{equation}\label{Bunch}
\upsilon_k\rightarrow\frac{e^{-i c_{s}k\eta}}{\sqrt{2c_s k}}.
\end{equation}
Subsequent to attain the numerical solutions of the MS equation (\ref{MS}), the accurate curvature fluctuations  power spectrum for  each mode $\upsilon_k$ can be acquired as
\begin{equation}\label{PsBunch}
{\cal P}_{\cal R}=\frac{k^3}{2\pi^2}\Big|{\frac{{\upsilon_k}^2}{z^2}}\Big|_{c_{s}k\ll aH}.
\end{equation}
Table \ref{tab2} embodies the  numerical outcomes for the precise  acme value of  the scalar power spectrum ${\cal P}_{\cal R}^{\rm peak}$, and associated  comoving wavenumber $k_{\rm peak}$ for each Case of Table \ref{tab1}. Moreover, in Fig. \ref{fig-ps} the precise power spectra for the entire Cases of  Table \ref{tab1} with regard to the  comoving wavenumber $k$, thereto the present observational confinements are delineated. In this figure the accurate   ${\cal P}_{\cal R}$  pertinent to the Cases A, B, C, and D portrays as the purple, green, red, and blue lines, respectively. Into the bargain, it can be inferred from the Fig. \ref{fig-ps} that, the amplitude of scalar power spectra for the entire Cases take nearly constant value around  ${\cal O}(10^{-9})$
in the course of slow-roll inflationary era on large scales in the environs of  the CMB scale ($k\sim0.05~ \rm Mpc^{-1}$), in consistency  with the present  observational data (\ref{psrestriction}). Whereas, intensification in the amplitude of power spectra to  order  ${\cal O}(10^{-2})$  during the USR era  on smaller scales can be perceived, which is adequate to produce detectable PBHs.
%===================================PBH Mass Fraction==================================================
\section{Generation of Primordial black holes}\label{sec5}
Subsequent to our preceding explications, this section is devoted to inquiry about formation of PBHs in NMDC framework emanated from sufficient multiplication of the amplitude of  curvature fluctuations on small scales.
Whereupon the super-horizon  enhanced fluctuation modes created during the inflationary era revert to the  horizon in the  RD epoch, the collapse of  ultra-condensed districts pertinent to these modes gives rise to generate PBHs.
The mass of nascent PBHs is pertaining to the horizon mass at the time of reverting by way of
\begin{align}\label{Mpbheq}
M_{\rm PBH}(k)=\gamma\frac{4\pi}{H}\Big|_{c_{s}k=aH} \simeq M_{\odot} \left(\frac{\gamma}{0.2} \right) \left(\frac{10.75}{g_{*}} \right)^{\frac{1}{6}} \left(\frac{k}{1.9\times 10^{6}\rm Mpc^{-1}} \right)^{-2},
\end{align}
where $\gamma$ denotes the efficiency of collapse, which is contemplated as  $\gamma=(\frac{1}{\sqrt{3}})^{3}$ \cite{carr:1975}, and $g_{*}$ signifies the efficient number of relativistic species upon thermalization in RD era, which is specified as  $g_{*}=106.75$  in the Standard Model of particle physics at high temperature.
Utilizing the Press-Schechter theory and with the presupposition of Gaussian statistics for distribution of curvature fluctuations, the formation rate for PBHs with mass $M(k)$ is computed \cite{Tada:2019,young:2014}  as the  subsequent form
\begin{equation}\label{betta}
  \beta(M)=\int_{\delta_{c}}\frac{{\rm d}\delta}{\sqrt{2\pi\sigma^{2}(M)}}e^{-\frac{\delta^{2}}{2\sigma^{2}(M)}}=\frac{1}{2}~ {\rm erfc}\left(\frac{\delta_{c}}{\sqrt{2\sigma^{2}(M)}}\right),
\end{equation}
wherein ``erfc" denotes the error function complementary, and $\delta_{c}$ depicts the threshold value of the density perturbations for PBHs production which is taken as $\delta_{c}=0.4$ pursuant to \cite{Musco:2013,Harada:2013}. Furthermore $\sigma^{2}(M)$ designates the coarse-grained density contrast with the smoothing scale $k$, and defines  as
\begin{equation}\label{sigma}
\sigma_{k}^{2}=\left(\frac{4}{9} \right)^{2} \int \frac{{\rm d}q}{q} W^{2}(q/k)(q/k)^{4} {\cal P}_{\cal R}(q),
\end{equation}
where ${\cal P}_{\cal R}$ is the curvature power spectrum, and  $W$  signifies  the window function which stipulated as  Gaussian window $W(x)=\exp{\left(-x^{2}/2 \right)}$.
In pursuance of determining  the  abundance of PBHs,
the present fraction of density parameters  related to  PBHs  $(\Omega_{\rm {PBH}})$ and Dark Matter $(\Omega_{\rm{DM}})$ is given as the subsequent form
\begin{equation}\label{fPBH}
f_{\rm{PBH}}(M)\simeq \frac{\Omega_{\rm {PBH}}}{\Omega_{\rm{DM}}}= \frac{\beta(M)}{1.84\times10^{-8}}\left(\frac{\gamma}{0.2}\right)^{3/2}\left(\frac{g_*}{10.75}\right)^{-1/4}
\left(\frac{0.12}{\Omega_{\rm{DM}}h^2}\right)
\left(\frac{M}{M_{\odot}}\right)^{-1/2},
\end{equation}
\begin{figure}[H]
\centering
\includegraphics[scale=0.6]{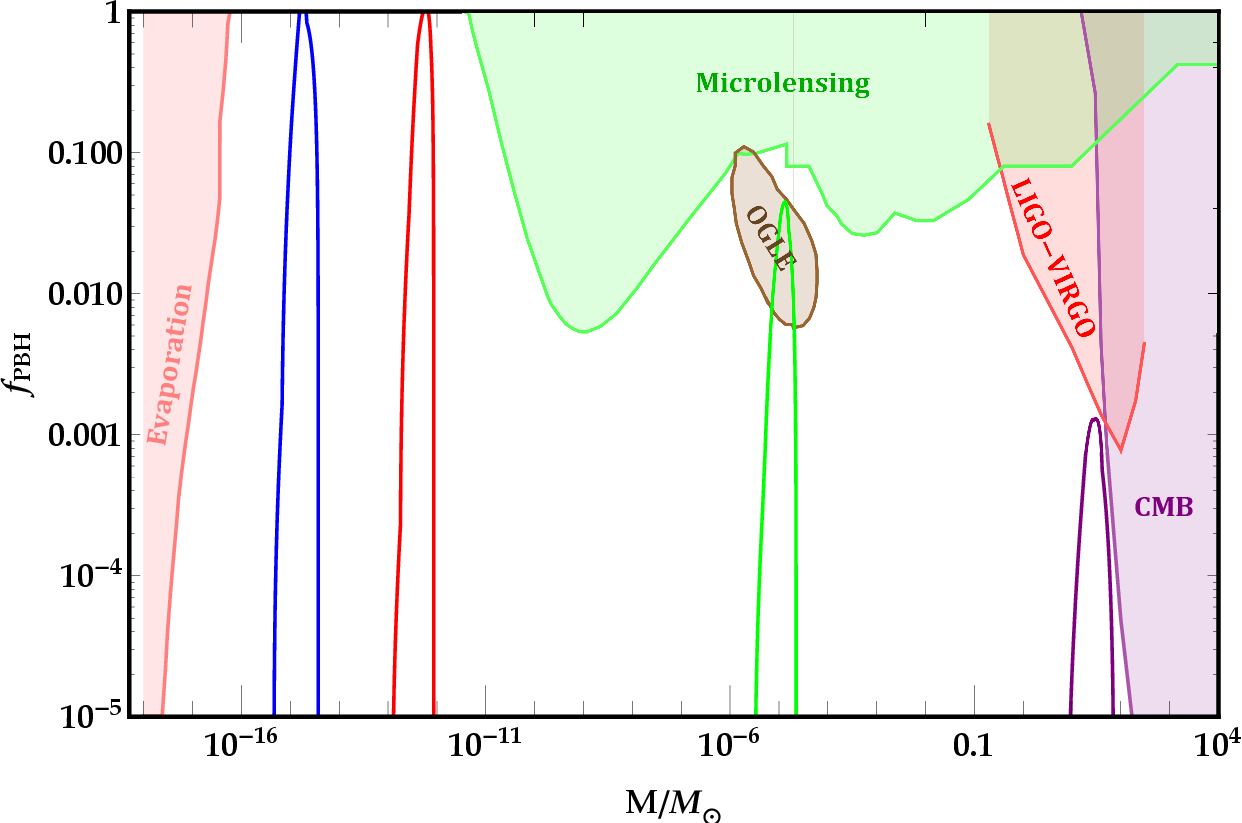}
\caption{The PBHs abundance $f_{\rm PBH}$ with regard to PBHs mass $M$ for the Cases A, B, C, and D delineated with  purple, green, red, and blue lines, respectively.  The shadowy domains illustrate the recent observational constraints on the fractional abundance of PBHs. The purple area depicts the restriction on CMB from signature of spherical accretion of PBHs inside  halos  \cite{CMB}.
The border of the red shaded domain depicts the upper bound  on the PBH abundance ensued from the LIGO-VIRGO event consolidation rate \cite{Abbott:2019,Chen:2020,Boehm:2021,Kavanagh:2018}. The brown shadowy domain portrays the authorized region for PBH abundance owing to  the ultrashort-timescale microlensing events in the OGLE data \cite{OGLE}. The green shaded area allots to constraints of  microlensing events from  cooperation between MACHO \cite{MACHO}, EROS \cite{EORS}, Kepler \cite{Kepler}), Icarus \cite{Icarus}, OGLE \cite{OGLE},  and Subaru-HSC \cite{subaro}. The pink shadowy region delineates the constraints related to PBHs evaporation such as
extragalactic $\gamma$-ray background \cite{EGG},  galactic center 511 keV $\gamma$-ray line (INTEGRAL) \citep{Laha:2019}, and effects on CMB spectrum \cite{Clark}.}
\label{fig-fpbh}
\end{figure}
\noindent
wherein the current density parameter of Dark Matter $\Omega_{\rm {DM}}h^2\simeq0.12$ is delineated by Planck 2018 data \cite{akrami:2018}.

Ultimately, by settling  the accurate scalar power spectrum acquired from numerical solution of Mukhanov-Sasaki equation in (\ref{sigma}) and utilizing  Eqs. (\ref{Mpbheq})-(\ref{fPBH}), we could compute the PBHs abundance for the entire  Cases of Table \ref{tab1}. Table \ref{tab2} and Fig. \ref{fig-fpbh} are delineated the attained numerical and plotted  upshots.

As a consequence of our finding, for parameter collection of Case A our model prognosticates PBHs with stellar-mass around $19.10M_{\odot}$ and abundance acme at $f_{\rm PBH}^{\rm peak}\simeq0.0012$, which are consonant with the upper bound  of the LIGO-VIRGO consolidation rate, and they can be  appropriate entrant to elucidate the GWs and the LIGO-VIRGO events.

Apropos of Case B the prognosticated PBHs with  earth-mass  around $M_{\rm PBH}^{\rm peak}=7.28\times10^{-6}M_{\odot}$ and abundance of $f_{\rm PBH}^{\rm peak}=0.0355$,  are localized  in the authorized  domain  of the ultrashort-timescale microlensing events in OGLE data, hence this Case of  PBHs  could be practical  to narrate microlensing events.

The parameter collections of Cases C and D  of our model, give rise to foretell two  PBHs mass spectra  in asteroid-mass range with masses $M_{\rm PBH}^{\rm peak}=2.713\times10^{-13}M_{\odot}$ and $M_{\rm PBH}^{\rm peak}=1.023\times10^{-15}M_{\odot}$, and acmes of $f_{\rm PBH}^{\rm peak}$ at around  $0.9615$ and $0.9526$. Ergo, they  could be contemplated as desirable nominee for the entire of  DM content.

It is worth noting that, recently the significant effects of quantum diffusion on curvature perturbations produced during USR, and  undeniable consequences of that on PBHs formation have been studied in literatures \cite{Ezquiaga,Pattison,Biagetti,Ballesteros,Pattison:2021,Figueroa:2021,Figueroa:letter2021,De,Ezquiaga:2018}. In \cite{Figueroa:2021,Figueroa:letter2021} applying the stochastic-$\delta N$ formalism in USR stage, due to the obtained exponential tail for distribution function of curvature perturbations, an increase about several orders of magnitude in the PBHs abundance in comparison with standard results has been computed.  However, in \cite{Cruces:2019,Firouzjahi:2019,Ballesteros:2020,Cruces:2022}  the conflicting results have been demonstrated. in \cite{Cruces:2019} it has been proven that, the stochastic effects during USR domain could be neglected due to invalidity of  $\delta N$ formalism and  separate universe approach in USR. On the other side the authors of \cite{Firouzjahi:2019} have confirmed the results of \cite{Cruces:2019} about the insignificancy of the stochastic effects in USR stage, but not because of the same reason. They have reproduced precisely the known leading classical donation of observable  power spectrum and bispectrum, applying the stochastic $\delta N$ formalism in the USR era. Subsequently  in \cite{Ballesteros:2020} it has been proven that, the concluded curvature power spectrum from stochastic inflation accurately fits, at the linear level, the numerical result computed of solving the Mukhanov-Sasaki equation, even in the USR phase. Moreover, the authors of \cite{Cruces:2022} have inferred that, extra information from the stochastic approach to inflation could not be attained in comparison with traditional perturbation theory, and there are no quantum diffusion dominated regimes in SR/USR inflation or even in the transition between these eras. Ergo, it is obvious that the effects of quantum diffusion in USR stage are currently under dispute in literatures, and we did not consider that in our model.
%====================================Induced Gravitational waves=========================================
\section{Produced GWs in NMDC Framework}\label{sec6}
\begin{figure}[H]
\centering
\includegraphics[scale=0.7]{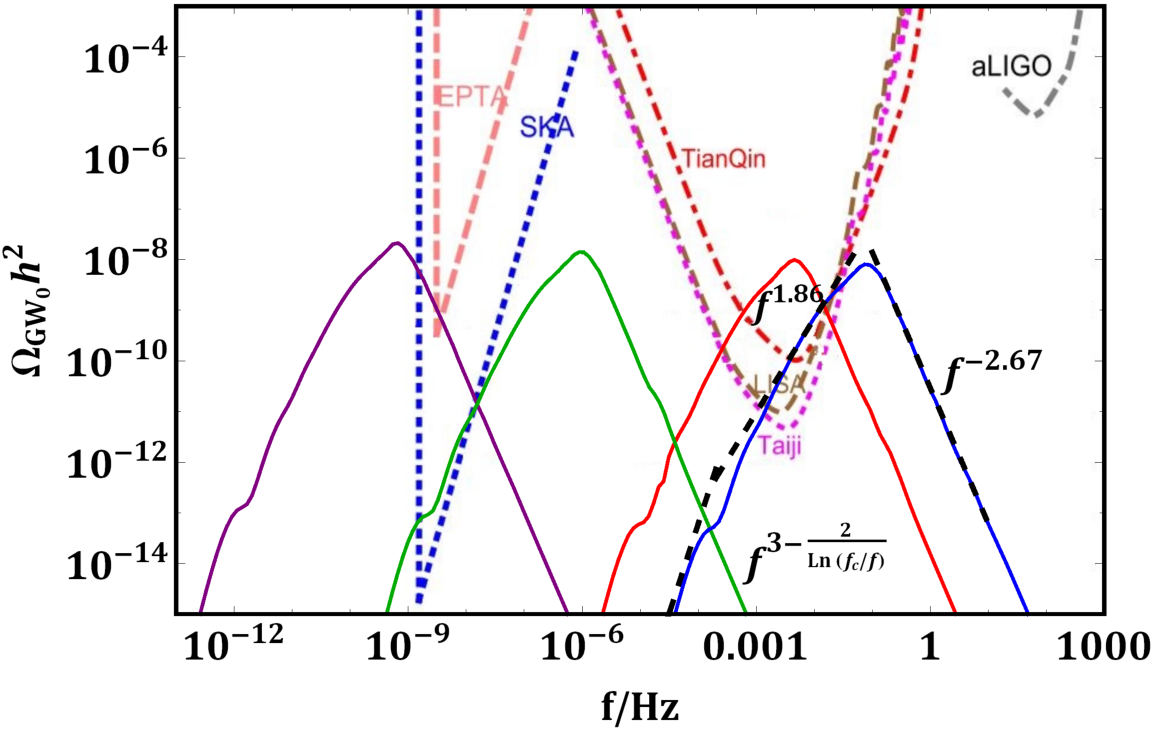}
\vspace{-0.5em}
\caption{ The procured spectra of present induced GWs energy density parameter $\Omega_{\rm GW_0}$ with regard to frequency pertinent to the Cases A (purple solid line), B (green solid line), C (red solid line), and D (blue solid line) of Table \ref{tab1}.  The  power-law behavior  of $\Omega_{\rm GW_0}$ is depicted  by black dashed line for the Case D.}
\label{fig-omega}
\end{figure}

Producing of induced GWs can be contemplated as another consequence of reverting the scales pertinent to enhanced amplitude of incipient curvature fluctuations to the horizon contemporaneous with generating of PBHs in RD era. The propagated GWs in the cosmos could be detected by dint of multifarious detectors if their energy spectra lie inside the sensitivity scopes of them. With this feature in mind, in this section we ponder how
the induced GWs in our NMDC model with exponential form of potential and coupling parameter can be produced.
It is substantiated that, with regard to  the second order impressions  of  perturbations theory, the mutation of tensor perturbations is emanated from the first order scalar perturbations.  The perturbed FRW metric utilizing  the conformal Newtonian gauge can be written as  the subsequent  form \cite{Ananda:2007}
\begin{eqnarray}
ds^2=a(\eta)^2\left\{-(1+2\Psi)d\eta^2 +\left[(1-2\Psi)\delta_{ij}+\frac{h_{ij}}{2}\right]dx^idx^j  \right\}\;,
\end{eqnarray}
wherein $\eta$, $\Psi$, and $h_{ij}$ intimate the conformal time, the first-order scalar perturbations, and the perturbation of the second-order transverse-traceless tensor, respectively.  Inasmuch as the inflaton field crumbles since  the end of inflationary epoch and transforms into light particles to heat the cosmos  to inaugurate the RD epoch, thus  the impact of inflaton field in the course of  cosmic mutation in RD era can be  disregarded. Thereupon,  the standard Einstein formulation can be utilized to inspect the production of induced GWs in RD era coeval with PBHs generation. In pursuance of this objective, the subsequent equation of motion is considered for the second-order tensor perturbations $h_{ij}$  \cite{Ananda:2007,Baumann:2007}
\begin{eqnarray}\label{EOM_GW}
h_{ij}^{\prime\prime}+2\mathcal{H}h_{ij}^\prime - \nabla^2 h_{ij}=-4\mathcal{T}^{lm}_{ij}S_{lm}\;,
\end{eqnarray}
in which   $\mathcal{H}\equiv a^{\prime}/a$   and  $\mathcal{T}^{lm}_{ij}$,  symbolize  the conformal Hubble parameter and  the transverse-traceless projection operator. The GW origin term $S_{ij}$ is contemplated  as
\begin{eqnarray}
S_{ij}=4\Psi\partial_i\partial_j\Psi+2\partial_i\Psi\partial_j\Psi-\frac{1}{\mathcal{H}^2}\partial_i(\mathcal{H}\Psi+\Psi^\prime)\partial_j(\mathcal{H}\Psi+\Psi^\prime)\; .
\end{eqnarray}
In the following the  scalar metric perturbations $\Psi$  in the Fourier space  in the course of  RD era is calculated  by  \cite{Baumann:2007} as
\begin{eqnarray}
\Psi_k(\eta)=\psi_k\frac{9}{(k\eta)^2}\left(\frac{\sin(k\eta/\sqrt{3})}{k\eta/\sqrt{3}}-\cos(k\eta/\sqrt{3}) \right)\;,
\end{eqnarray}
wherein $k$  signifies the comoving wavenumber. Furthermore the incipient perturbations $\psi_k$ is affiliated to  the curvature fluctuations power spectrum  by dint of  the following  two-pointed correlation function
\begin{eqnarray}
\langle \psi_{\bf k}\psi_{ \tilde{\bf k}}  \rangle = \frac{2\pi^2}{k^3}\left(\frac{4}{9}\mathcal{P}_{\cal R}(k)\right)\delta(\bf{k}+ \tilde{\bf k})\;.
\end{eqnarray}
In the end, the following equation for energy density of induced GWs in the course of RD era is attained by \cite{Kohri:2018}
\begin{eqnarray}\label{OGW}
&\Omega_{\rm{GW}}(\eta_c,k) = \frac{1}{12} {\displaystyle \int^\infty_0 dv \int^{|1+v|}_{|1-v|}du } \left( \frac{4v^2-(1+v^2-u^2)^2}{4uv}\right)^2\mathcal{P}_{\cal R}(ku)\mathcal{P}_{\cal R}(kv)\left( \frac{3}{4u^3v^3}\right)^2 (u^2+v^2-3)^2\nonumber\\
&\times \left\{\left[-4uv+(u^2+v^2-3) \ln\left| \frac{3-(u+v)^2}{3-(u-v)^2}\right| \right]^2  + \pi^2(u^2+v^2-3)^2\Theta(v+u-\sqrt{3})\right\}\;,
\end{eqnarray}
in which  $\Theta$ denotes the Heaviside theta function, and $\eta_{c}$ designates the time of ceasing the growth of   $\Omega_{\rm{GW}}$.
By way of the subsequent equation, the present value of  the induced GWs energy spectra  is affiliated to the energy spectra at $\eta_{c}$  \cite{Inomata:2019-a}
\begin{eqnarray}\label{OGW0}
\Omega_{\rm GW_0}h^2 = 0.83\left( \frac{g_{*}}{10.75} \right)^{-1/3}\Omega_{\rm r_0}h^2\Omega_{\rm{GW}}(\eta_c,k)\;,
\end{eqnarray}
in which the present value of radiation density parameter is indicated by $\Omega_{\rm r_0}h^2\simeq 4.2\times 10^{-5}$, and $g_{*}\simeq106.75$ signifies  the effective degrees of freedom in the energy density at $\eta_c$. Moreover frequency is related to wavenumber through the following equation
\begin{eqnarray}\label{k_to_f}
f=1.546 \times 10^{-15}\left( \frac{k}{{\rm Mpc}^{-1}}\right){\rm Hz}.
\end{eqnarray}
In follow up our study, we utilize the accurate scalar power spectrum deduced from numerical solution of the MS equation beside Eqs. (\ref{OGW})-(\ref{k_to_f}), and could  acquire  the present energy  spectra of  scalar induced GWs associated with PBHs for the entire  Cases of Table \ref{tab1}. The diagram of our foretold upshots beside the susceptibility curves of varietal GWs observatories are depicted in Fig. \ref{fig-omega}. It is worth noting that, rectitude  of our prognosticated  upshots can be verified  in view of these  GWs observatories which are composed of  European PTA (EPTA) \cite{EPTA-a,EPTA-b,EPTA-c,EPTA-d}, the Square Kilometer Array (SKA)  \cite{ska}, Advanced Laser Interferometer
\noindent
Gravitational Wave Observatory (aLIGO) \cite{ligo-a,ligo-b}, Laser Interferometer Space Antenna (LISA)  \cite{lisa,lisa-a}, TaiJi \cite{taiji}, and TianQin  \cite{tianqin}.

It is obvious from Fig. \ref{fig-omega} that, the acmes of  spectra of $\Omega_{\rm GW0}$ prognosticated  from our setup, for the Cases A (purple line), B (green line), C (red line), and D (blue line)  have located  at disparate frequencies  with approximately alike  altitude  of order $10^{-8}$.
For the Cases A and B appertain to  stellar and earth mass PBHs, the acmes of spectra of $\Omega_{\rm GW_0}$ have localized at frequencies around  $f_{c}\sim10^{-10}\text{Hz}$ and $f_{c}\sim10^{-7}\text{Hz}$ respectively, which lie inside the susceptibility scope of the  SKA observatory. Apropos of the Cases C and D pertinent to asteroid mass PBHs,  the acmes of $\Omega_{\rm GW_0}$ spectra     settle in  mHz and cHz frequency zone which could be traced by LISA, TaiJi, and TianQin observatories. Inasmuch as,  the prognosticated spectra of $\Omega_{\rm GW0}$ for the entire Cases of our model  could cross the susceptibility curves of disparate GWs observatories,  the  legitimacy  of this model could be appraised by way of  the broadcasted  data of these observatories in future.

After all we  scrutinize the slope of spectra of $\Omega_{\rm GW_0}$ at the disparate frequency zones in the environs of the acme position. Newly, it has been corroborated that in the vicinity  of acme position, the current density parameter spectra of induced GWs have a power-law behaviour with regard  to frequency as  $\Omega_{\rm GW_0} (f) \sim f^{n} $ \cite{fu:2020,Xu,Kuroyanagi}. In Fig. \ref{fig-omega},  the approximate slope  of the spectrum of $\Omega_{\rm GW_0}$ are  mapped  with black dashed lines in three ranges of frequency for the Case D. For this Case we have computed the frequency of acme as $f_{c}=0.0744{\rm Hz}$,  and appraise the power index of power-law function  $n=1.86$ for $f<f_{c}$,  $n=-2.67$ for $f>f_{c}$, and  moreover a log-reliant form as $n=3-2/\ln(f_c/f)$ for the infrared region $f\ll f_{c}$,  which is consonant with the analytic sequels procured in \cite{Yuan:2020,shipi:2020}.
%========================================Conclusions=====================================================
\section{Conclusions}\label{sec7}
In this work, the PBHs generation in the  inflationary model  pertinent  to the Horndenski theory with nonminimal  coupling between the field derivative and the Einstein tensor expounded by action (\ref{action}) is verified.
The enhancement of friction gravitationally emanated from nonminimal field derivative coupling to gravity setup making  the inflaton slow down and  gives rise to a transient stage in the dynamics of scalar field namely ultra slow-roll inflationary era.
With this trait in mind, whereas in standard framework of inflation the exponential potential drives an endless inflationary epoch \cite{karami:2017} and, as regards the inconsistency of the prognostications of this form of potential at CMB scales  with Planck  2018 data \cite{akrami:2018,karami:2017}, we contemplated it  in NMDC setup and try to amend its foretold outcomes.

By  considering exponential  potential for our model beside defining coupling parameter as  two-parted exponential function of inflaton field (\ref{t})-(\ref{tII}), and thence  fine-tuning of the four parameter collections (A, B, C, and D) depicted in Table \ref{tab1},  we were able to slow down the inflaton velocity adequately to produce PBHs in an ultra slow-roll phase on  small scales. Another consequence of these choices is that the observational results  of the model were obtained in accordance with Planck 2018 data on CMB scales.

Furthermore, we delineated mutation diagram of inflaton field $\phi$, the first and second slow-roll parameters ($\varepsilon$ and $\delta_{\phi}$) in terms of  $e$-fold number $N$ in Figs. \ref{fig:phi} and \ref{fig:SRp} by way of accurate solving of the  background equations  (\ref{FR1:eq})-(\ref{Field:eq}).
As regards the mutation diagram of slow-roll parameters in Fig. \ref{fig:SRp} it can be inferred that, in the course of USR stage $\varepsilon$  adheres to the slow-roll provisos ($\varepsilon\ll1$) but $\delta_{\phi}$ contravenes that ($\left|\delta_{\phi}\right|\gtrsim1$). Hence, we calculated the accurate curvature fluctuations power spectra appertain to the entire Cases of Table \ref{tab1} by numerical solving of the  Mukhanov-Sasaki equation. The numerical upshots epitomized in Table \ref{tab2} and schemed diagram in  Fig. \ref{fig-ps} illustrate that, the attained accurate power spectra have approximately constant values in consistency  with the Planck 2018 data on CMB scales, whereas on smaller  scales they have acmes with  adequate altitude to produce detectable PBHs.

Regarding the obtained numerical results for $n_{s}$ and $r$ enumerated in Table \ref{tab2},  we can see  that in view of Planck  2018  TT,TE,EE+lowE+lensing+BK14+BAO data the quantity of
$r$ for the entire Cases and $n_{s}$ for the Cases B, C, and D are consonant  with $68\%$ CL, whereas $n_{s }$ for the Case A is  consonant with $95\%$  CL \cite{akrami:2018}. As a consequence of choosing NMDC groundwork and appropriate exponential form of coupling parameter for our model, we could rectify the observational prognostications of exponential potential.

At length by utilizing   the accurate scalar power spectrum acquired from numerical solution of Mukhanov-Sasaki equation and  Press-Schechter formulation, we could compute the PBHs abundance for the entire  Cases of Table \ref{tab1}.
Prognosticated PBHs for the Case A  with stellar-mass around $19.1~M_{\odot}$ and abundance acme at $f_{\rm PBH}^{\rm peak}\simeq0.0012$  could  be  appropriate  to explicate the GWs and the LIGO-VIRGO events. Obtained PBHs for the Case B with  earth-mass around $M_{\rm PBH}^{\rm peak}=7.28\times10^{-6}M_{\odot}$ and abundance of $f_{\rm PBH}^{\rm peak}=0.0355$ could be practical to narrate microlensing events. Moreover foretold PBHs for the Cases C and D with masses $M_{\rm PBH}^{\rm peak}=2.713\times10^{-13}M_{\odot}$ and $M_{\rm PBH}^{\rm peak}=1.023\times10^{-15}M_{\odot}$, and acmes of $f_{\rm PBH}^{\rm peak}$ at around  $0.9615$ and $0.9526$ could be contemplated as desirable nominee for the entire of DM content (see Table \ref{tab2} and Fig. \ref{fig-fpbh}).

At last, we investigated  production  of the induced GWs coeval with   PBHs formation and computed the current spectra of $\Omega_{\rm GW_0}$ for the entire Cases of our model. It is inferred from our computation that, all Cases have located  at disparate frequencies  with approximately alike  altitude   of order $10^{-8}$ (see Fig. \ref{fig-omega}). For the Cases A and B appertain to  stellar and earth mass PBHs, the acmes of  $\Omega_{\rm GW_0}$ spectra have localized at frequencies around  $f_{c}\sim10^{-10}\text{Hz}$ and $f_{c}\sim10^{-7}\text{Hz}$ respectively, which lie inside the susceptibility scope of the  SKA observatory. Apropos of the Cases C and D pertinent to asteroid mass PBHs,  the acmes of $\Omega_{\rm GW_0}$ spectra settle in  mHz and cHz frequency zone which could be traced by LISA, TaiJi, and TianQin observatories. Ergo the legitimacy  of our  model could be appraised by way of the broadcasted  data of these observatories in future.

After all we  checked the power-law behaviour of  the slope of spectra of $\Omega_{\rm GW_0}$ at the disparate frequency zones in the environs of the acme position with regard  to frequency as  $\Omega_{\rm GW_0} (f) \sim f^{n} $ \cite{fu:2020,Xu,Kuroyanagi}, and we appraised the power index of power-law function  $n=1.86$ for $f<f_{c}$,  $n=-2.67$ for $f>f_{c}$, and  moreover a log-reliant form as $n=3-2/\ln(f_c/f)$ for the infrared region $f\ll f_{c}$,  which is consonant with the analytic sequels of \cite{Yuan:2020,shipi:2020}.
%===========================================Refrence======================================================
%\newpage

\end{document}